\documentclass[11pt,a4paper]{article}

\usepackage{jstyle}

\author[a]{Minkyeu CHO}
\author[a]{\quad Euihun JOUNG}
\author[b]{\quad TaeHwan OH}
\author[a]{\quad Tung TRAN}

\affiliation[a]{Department of Physics, College of Sciences, and Research Institute of Basic
  Science, \\ Kyung Hee University, Seoul 02447, Korea}
\affiliation[b]{%
    Quantum Universe Center, \\
    Korea Institute for Advanced Study,
   Seoul 02455, Korea.
}

\emailAdd{minku8542@khu.ac.kr}
\emailAdd{euihun.joung@khu.ac.kr}
\emailAdd{hepthoh@kias.re.kr}
\emailAdd{tung.tran@khu.ac.kr}

\title{Worldline Higher Spin Gravity}

\abstract{
We propose a worldline formulation of higher-spin gravity (HSG) in $\mathrm{AdS}_4$, based on a simple twistor action. Taken at face value, the model describes only the free propagation of massless higher-spin fields. The central observation of this work is that the model admits a natural \emph{double-line} interpretation, which supplies a geometric prescription for gluing worldlines at interaction vertices, in close parallel with the joining of strings in string theory. 
Building on this picture, we construct $\mathrm{AdS}$-covariant vertex operators for all massless higher-spin fields, show that they satisfy the Bargmann--Wigner equations, and use them to compute the $n$-point correlation functions of type-A and type-B HSG as worldline path integrals of these vertex operators.
In the boundary limit these correlators reproduce the higher-spin current correlators of free boson and free fermion vector models. We further discuss the embedding of the worldline theory into Poisson sigma model 
, where the doubled-line structure acquires a geometric origin as the two edges of an open string worldsheet, together with several consequences of this enlarged framework---fractional branes, loop expansion, unoriented projection, and the prospect of a worldsheet formulation of HSG.
}

\begin{document}

\maketitle

\section{Introduction}

Vasiliev's higher-spin gravity \cite{Vasiliev:1990en,Vasiliev:2003ev} has long been regarded as a toy model for the tensionless limit of string theory. Like string theory, it contains an infinite tower of higher-spin states; unlike string theory, however, all of these states are massless, precisely as expected in the tensionless regime. 
Despite this infinite spectrum, the field content of higher-spin gravity is in fact drastically simpler than that of string theory: it consists solely of totally symmetric massless fields, each appearing with multiplicity one, and can therefore be understood as the tensionless limit of the first Regge trajectory---the spectrum generated by the string oscillator $\a_{-1}^\mu$ \cite{Bengtsson:1986ys,Pashnev:1997rm,Sagnotti:2003qa}; see also \cite{Polyakov:2009pk,Eberhardt:2019ywk} for other attempts to derive theories of massless higher-spin fields from string theory.

Higher-spin gravity (HSG) is governed by Vasiliev's equations, while a corresponding action principle remains elusive.\footnote{See, e.g., \cite{Vasiliev:2005zu,Boulanger:2011dd,Boulanger:2012bj,Arias:2016ajh,Tarusov:2021fdk} for previous attempts to construct an action for higher-spin gravity.} Structurally, these equations can be viewed as a generalization of Cartan's formulation of gravity within the framework of free differential algebras \cite{Sullivan:1977pdi,vanNieuwenhuizen:1982zf,
Vasiliev:1988xc}, in which the role of the spacetime isometry algebra is subsumed into a larger algebra---the higher-spin algebra. A well-known feature of Vasiliev's equations is that they involve infinitely many auxiliary fields, which renders basic field-theoretic questions such as locality highly nontrivial\footnote{Following the observation \cite{Bekaert:2014cea,Vasiliev:2015wma,Boulanger:2015ova,Bekaert:2015tva,Skvortsov:2015lja} of a subtle interplay between non-locality and field redefinitions in Vasiliev's higher-spin gravity, significant efforts have been devoted to this problem; see, e.g., \cite{Didenko:2015cwv,Vasiliev:2017cae,Didenko:2018fgx,Didenko:2019xzz} and the very recent work \cite{Kirakosiants:2025gpd} and references therein.}---a characteristic reminiscent of string field theory.

A central reason for the conceptual power of string theory is that its fundamental ingredient is the string itself: the worldsheet description provides an organizing principle from which all target-space interactions can be derived. In view of the structural parallels between higher-spin gravity and string theory, it is natural to ask whether HSG admits a comparable worldsheet---or, at a more modest level, worldline---formulation. A worldsheet-like reformulation of HSG would supply an organizing principle from which the intricate non-localities of the Vasiliev system might emerge in a controlled fashion.

The relationship between HSG and string theory can in fact be approached from two complementary angles. Viewed as a theory of gravity, HSG arises naturally as the tensionless limit of closed string theory, with the graviton accompanied by an infinite tower of massless higher-spin partners. Viewed instead from the perspective of Chan--Paton dressing \cite{Konstein:1989ij, Vasiliev:2004cm},\footnote{E.J. thanks A. Sagnotti for emphasizing this point.} an analogous picture arises on the open-string side, with the higher-spin fields interpreted as the massless limit of their massive counterparts on the first Regge trajectory. Since the natural backgrounds for HSG are anti-de Sitter (AdS) and de Sitter (dS) spacetimes, one may envision a tensionless string whose characteristic length is comparable to the (A)dS radius, and whose endpoints reach the asymptotic boundary.

Despite the appeal of this direction, very few attempts have been pursued along this line. The most notable is the proposal of Engquist and Sundell \cite{Engquist:2005yt}, who argued that higher-spin gravity, tensionless strings, and branes in AdS all admit a common description in terms of so-called \emph{partonic singletons}---boundary degrees of freedom on which strings and higher-dimensional objects may end. In this picture, a string or brane is built from $N$ elementary constituents, the partons, each of which lives on the boundary of AdS and behaves as a free singleton. The extended object is then recovered in the large-$N$ limit: as $N\to\infty$, the symmetric collective state of $N$ singletons is conjectured to develop an emergent worldsheet together with a Chan--Paton structure, with the worldsheet itself arising geometrically from the collective dynamics of the partons. From a representation-theoretic standpoint, this construction generalizes the Flato--Fronsdal theorem \cite{Flato:1978qz}: rather than tensoring two singletons to obtain the tower of massless higher-spin fields, one symmetrically tensors arbitrarily many---see, e.g., \cite{Bae:2016rgm}---to generate extended objects such as strings.

The worldsheet theory of strings may be illuminated by drawing an analogy with the worldline formulation of a relativistic particle. This idea is supported by a substantial body of work on the worldline side, in which the formulation has been generalized to particles with spin, typically through the introduction of additional spin variables together with a suitable set of constraints \cite{Gershun:1979fb,Howe:1988ft,Kuzenko:1994ju,Lyakhovich:1996we,Lyakhovich:1998ij,Bastianelli:2007pv,Bastianelli:2008nm,Bastianelli:2014lia,Kuzenko:2020ayk,Andrzejewski:2020qxt,Basile:2023vyg}. 
Both worldsheet string theory 
and these worldline theories share the use of position variables, which lends them an intuitive physical interpretation. By contrast, the twistor formulations of worldline particles \cite{Shirafuji:1983zd,Howe:1992bv,Claus:1999zh,
Zunger:2000wy,Cederwall:2000km,Fedoruk:2014vqa,Mezincescu:2015apa,Arvanitakis:2016vnp,Arvanitakis:2016wdn,Arvanitakis:2017cpk,Uvarov:2019vmd,Joung:2024akb} and worldsheet strings \cite{Witten:2003nn,Berkovits:2004hg,Berkovits:2004jj} are  less transparent from this standpoint, but offer a substantial advantage in simplicity, typically involving fewer variables and constraints.

Of particular relevance to the present work is the twistor formulation of massless higher-spin fields in AdS$_4$, AdS$_5$, and AdS$_7$ \cite{Cederwall:2000km,Arvanitakis:2016vnp,Arvanitakis:2016wdn,Arvanitakis:2017cpk}, realized in terms of real, complex, and quaternionic twistors, respectively. These models are described by a one-derivative Hamiltonian action in which half of the twistor variables are position-like and the remaining half are momentum-like. 
The Hamiltonian itself consists entirely of constraints, enforced through Lagrange multipliers. Among these, one constraint is specifically responsible for fixing the spin of the underlying particle, taking the form
\be
\cC=\cS-s\approx 0\,,
\label{spin const}
\ee
where $\cC$ denotes the constraint imposed with its Lagrange multiplier partner, $\cS$ is a simple function of the twistor variables, and $s$ is the spin of the particle.

In \cite{Cederwall:2004cf}, Cederwall observed that, upon dropping the spin constraint \eqref{spin const} from the full set of constraints, the resulting model describes a particle spectrum identical to that of HSG. The case of AdS$_4$ is especially simple: here, \eqref{spin const} is in fact the only constraint present, so that removing it yields a remarkably simple unconstrained action with vanishing Hamiltonian,
\be
S=\int \O_{AB}\,\bm Z^{A}\cdot \dd \bm Z^{B}\,,
\label{eq:action-Z}
\ee
where $\O_{AB}$ is the symplectic form, $A, B$ are $Sp(4,\mathbb{R})$ vector indices, and $\bm Z^A$ itself carries two components $(Z_1^A, Z_2^A)$, with the inner product defined as $\bm Z^A\cdot \bm W^B = Z_1^A\,W_1^B + Z_2^A\,W_2^B$. This internal two-component structure encodes a dual group, on whose role we shall comment in due course. For other twistorial approaches to higher-spin particles in flat spacetime, we refer the reader to \cite{Fedoruk:2006it,Fedoruk:2007ud,Fedoruk:2005np,Fedoruk:2006jm}, in which the first quantization of the system leads to equations closely resembling those of the Vasiliev theory.

The present paper investigates the possibility of interpreting the action \eqref{eq:action-Z} as a worldline model for fully interacting HSG, by pursuing the analogy with the worldsheet theory of strings. Taken at face value, the model appears to describe only the free propagation of massless spinning fields, corresponding to the perturbative spectrum of HSG. Moreover, the worldline theory as such provides no a priori prescription for gluing the twistor variables at branching vertices:\footnote{See \cite{Neiman:2022enh} for a recent proposal of a worldline framework for HSG based on Feynman diagrammatics.}
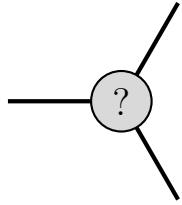
\begin{center}
\begin{tikzpicture}[thick]
\coordinate (O) at (0,0);
\draw[ultra thick] (O) -- (180:1.5);
\draw[ultra thick] (O) -- (60:1.5);
\draw[ultra thick] (O) -- (300:1.5);
\filldraw[fill=gray!30, draw=black] (O) circle (0.4);
\node at (O) {\Large ?};
\end{tikzpicture}
\end{center}
This situation, however, is closely analogous to the one familiar from string theory: the worldsheet action itself describes only free propagation, while the gluing prescription at interaction vertices is supplied as additional input, naturally interpreted as the smooth geometric joining of strings.

The central observation of this work is that the worldline model \eqref{eq:action-Z} naturally admits a \emph{double-line} interpretation, in which the two components $Z_1^A$ and $Z_2^A$ are regarded as living on two distinct lines. This double-line structure provides precisely the room needed to prescribe gluing conditions at each vertex in a geometric manner, in close parallel with string theory:
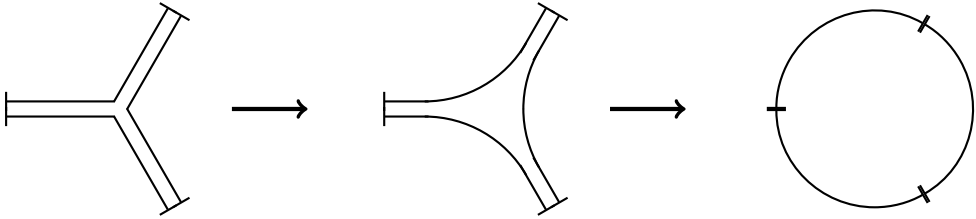
\begin{center}
\begin{tikzpicture}[thick]
\draw[shift={(150:0.1)},-|] (60:0.05) -- (60:1.5);
\draw[shift={(150:-0.1)},-|] (60:0.05) -- (60:1.5);
\draw[shift={(270:-0.1)},|-] (180:1.5) -- (180:.05);
\draw[shift={(270:0.1)},|-] (180:1.5) -- (180:.05);
\draw[shift={(-150:-0.1)},-|] (-60:0.05) -- (-60:1.5);
\draw[shift={(-150:0.1)},-|] (-60:0.05) -- (-60:1.5);
\draw[ultra thick,->] (1.5,0)--(2.5,0);
\begin{scope}[shift={(5,0)}]
\draw[shift={(150:0.1)},-|] (60:0.8) -- (60:1.5);
\draw[shift={(270:-0.1)}] (180:0.95) arc (-90:-30:1.55);
\draw[shift={(270:-0.1)},|-] (180:1.5) -- (180:.9);
\draw[shift={(270:0.1)},|-] (180:1.5) -- (180:.9);
\draw[shift={(270:0.1)}] (180:0.95) arc (90:30:1.55);
\draw[shift={(-150:0.1)},-|] (-60:0.8) -- (-60:1.5);
\draw[shift={(150:-0.1)},-|] (60:0.8) -- (60:1.5);
\draw[shift={(150:-0.1)}] (60:0.95) arc (150:180:1.55);
\draw[shift={(-150:-0.1)}] (-60:0.95) arc (210:180:1.55);
\draw[shift={(-150:-0.1)},-|] (-60:0.8) -- (-60:1.5);
\end{scope}
\draw[ultra thick,->] (6.5,0)--(7.5,0);
\begin{scope}[shift={(10,0)}]
\draw[|-|] (60:1.3) arc (60:180:1.3);
\draw[|-|] (180:1.3) arc (180:300:1.3);
\draw[|-|] (300:1.3) arc (300:420:1.3);
\end{scope}
\end{tikzpicture}
\end{center}
Pursuing the string-theoretic analogy further, we introduce a set of vertex operators compatible with the picture above. These operators are constructed in an AdS-covariant manner and are shown to satisfy the standard free equations of motion for massless fields in AdS. Using them, we compute the bulk $n$-point correlation functions of HSG as $n$-point correlators of vertex operators on the worldline. Their boundary limit, with a suitable rescaling, reproduces the correlation functions of the free boson and free fermion conformal field theories (CFTs), thereby confirming that the worldline model \eqref{eq:action-Z} is capable of describing both type-A and type-B HSG.\footnote{The holographic correspondence of HSG \cite{Sezgin:2002rt,Klebanov:2002ja,Sezgin:2003pt} states that the type-A and type-B HSG are dual, respectively, to the bosonic and fermionic vector model CFTs, with the boundary condition imposed on the bulk scalar selecting between the free and the critical theory. This correspondence has been verified by Giombi and Yin  for a large class of three-point functions \cite{Giombi:2009wh,Giombi:2010vg,Giombi:2012ms};
see \cite{Giombi:2016ejx} for a review.
} 
Our results also bear a strong resemblance to \cite{Colombo:2012jx,Didenko:2012tv,Didenko:2013bj}, which compute the star-product trace of bulk-to-boundary propagator zero-forms in Vasiliev theory.

We further note that the worldline along the split single line can be viewed as the parton action of \cite{Engquist:2005yt} specialized to the twistor space of AdS$_4$. As also remarked in \cite{Engquist:2005yt}, the worldline model for HSG can in turn be regarded as the boundary reduction of the Poisson sigma model (PSM) \cite{Ikeda:1993fh,Schaller:1994es}, where the boundary of the worldsheet is identified with the worldline.

\paragraph{Outline.}
The remainder of the paper is organized as follows.

In Section \ref{sec:2}, we study the worldline model \eqref{eq:action-Z} and describe its path integral. Using AdS$_4$ isometries, we construct AdS-covariant operators for the scalar field and for the higher-spin Weyl tensors, and verify that they satisfy the Klein--Gordon and Bargmann--Wigner equations, respectively. From these operators we compute the AdS$_4$ two-point functions and find that they contain contributions from reflection off the asymptotic boundary, which also flips the helicity of the propagating fields---an effect reminiscent of the method of images in electromagnetism.

In Section \ref{sec:3}, we introduce a prescription for gluing worldline fields together, inspired by the worldsheet theory of strings. Reinterpreting our model \eqref{eq:action-Z} as a worldline theory on a doubled line, we glue the two single lines at vertices and define, from each state, a vertex operator acting on a single line at the endpoint of the doubled line. This construction may be viewed as the worldline counterpart of the state--operator correspondence in two-dimensional CFT:
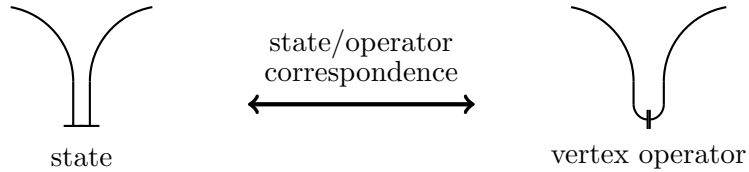
\begin{center}
\begin{tikzpicture}[thick]
\begin{scope}[shift={(1,0)}]
\draw[|-] (-0.11,-0.1) -- (-.11,0.5);
\draw (-.11,0.5) arc (0:80:1);
\draw[|-] (.11,-0.1) -- (.11,0.5);
\draw (.11,0.5) arc (180:100:1);
\node at (0,-0.5) {state};
\end{scope}
\node at (4.7,1) {state/operator};
\node at (4.7,0.6) {correspondence};
\draw[<->,ultra thick] (3.2,0.2)--(6.2,0.2);
\begin{scope}[shift={(8.5,0)}]
\draw (-.2,0.2) -- (-0.2,0.5);
\draw (.2,0.2) -- (0.2,0.5);
\draw (-.2,0.5) arc (0:80:1);
\draw (.2,0.5) arc (180:100:1);
\draw [-|] (-.2,0.2) arc (180:270:0.2);
\draw [-|] (.2,0.2) arc (360:270:0.2);
\node at (0,-0.5) {vertex operator};
\end{scope}
\end{tikzpicture}
\end{center}
The vertex operators are not uniquely fixed by this construction: they are defined only up to a linear combination of two possibilities. We resolve this ambiguity by imposing a superselection rule dictated by a $\mathbb{Z}_2$ quotient of the target twistor space, which selects precisely two distinct choices, corresponding to type-A and type-B HSG.

Section \ref{sec:4} specializes to the type-A and type-B theories. Using the vertex operators introduced above, we compute the bulk $n$-point correlation functions of higher-spin fields in AdS$_4$. Since the computation reduces to Gaussian integrals, exact results are obtained for arbitrary $n$. In the boundary limit, these reproduce the correlators of higher-spin currents in the three-dimensional free boson and free fermion CFTs. We conclude the section with a comparison to earlier works: the construction is formally analogous to the Vasiliev-based computations of \cite{Colombo:2012jx,Didenko:2012tv,Didenko:2013bj}---via an explicit oscillator dictionary---and conceptually aligned with the brane-parton proposal of \cite{Engquist:2005yt}, though differences remain in both cases.

Section \ref{sec:5} is devoted to a broader discussion of the framework. We first highlight the parallels with string theory. We then reformulate the worldline model as the edge mode of a PSM, an enlarged framework in which the double-line structure acquires a natural geometric origin and in which several features---fractional branes at orbifold fixed points, a loop expansion organized by worldsheet topology, and an unoriented projection with associated Chan--Paton structure---become considerably more transparent. We close with a number of future directions.

The paper is supplemented by three appendices to which various technical details are relegated. They summarize our notations and conventions, discuss the connection to Howe duality, and present the details of Gaussian integral and its expansion in the radial coordinates $z$.

%%%%%%%%%%%%%%%%%%%%%%%%%%%%%%%%%%%%%%%%%%%%
\section{Worldline model as free theory}\label{sec:2}
%%%%%%%%%%%%%%%%%%%%%%%%%%%%%%%%%%%%%%%%%%%%

In this section, we study the worldline model \eqref{eq:action-Z} as a free theory and identify its physical states with massless higher-spin fields in AdS$_4$.

The twistor variable $\bm Z^A=(Z_1^A, Z_2^A)$ in the action \eqref{eq:action-Z} decomposes into position- and momentum-like variables $\bm \xi^\a=(\xi_1{}^\a, \xi_2{}^\a)$ and $\bm \pi_\a=(\pi_{1\a}, \pi_{2\a})$ as
\be
	\bm Z^A=\binom{\bm \xi^\a}{\bm \pi_{\a}}
    =\begin{pmatrix}
        \xi_1{}^\a & \xi_2{}^\a \\
        \pi_{1\a} & \pi_{2\a}
    \end{pmatrix},
	\qquad \text{with}
	\qquad \a=1,2\,.
\ee
With the symplectic form
\be \label{symplectic form}
    \Omega_{AB}=\frac12\begin{pmatrix}0& -\delta_{\b}^\a\\ \delta_\a^\b & 0
\end{pmatrix},
\ee
and after integration by parts, the action \eqref{eq:action-Z} reduces to
\be \label{twistor action}
	S = -\int  \dd \bm\pi_{\a}\cdot \bm\xi^\a\,.
\ee
This simple action governs the transition amplitude between two states $|\Psi_1\ra$ and $|\Psi_2\ra$,
\begin{align}
    \la \Psi_2 | \Psi_1 \ra=
    \int D\bm Z\,
    \la\Psi_2|\bm \pi({\rm e}I)\ra
    \,\la\bm \pi({\rm s}I)|\Psi_1\ra\,
    \exp\left(-i\,\int_I\dd \bm\pi_\a\cdot \bm\xi^\a\right),
\end{align}
where ${\rm s}I$ and ${\rm e}I$ denote the starting and ending points of a worldline interval $I$, and $D\bm Z=D\bm\xi\,D\bm\pi$ is the path integral measure. The transition between $|\Psi_1\ra$ at ${\rm s}I$ and $|\Psi_2\ra$ at ${\rm e}I$, with time flowing from the starting to the ending point, may be depicted as follows:
\begin{center}
\begin{tikzpicture}[thick]
\begin{scope}
\draw[ultra thick] (0,0) -- (1.5,0);\draw[ultra thick,<-] (1.4,0) -- (3,0);
\node at (0.1,-0.3) {${\rm e}I$};
\node at (-0.7,0) {$\la\Psi_2|$};
\node at (2.9,-0.3) {${\rm s}I$};
\node at (3.7,0) {$|\Psi_1\ra$};
\node at (1.5,0.4) {$I$};
\end{scope}
\end{tikzpicture}
\end{center}
Since the worldline action \eqref{twistor action} is first-order in derivatives, a boundary condition may be imposed on either $\bm\xi^\a$ or $\bm\pi_\a$ in $\bm Z^A=(\bm\xi^\a, \bm\pi_\a)$. We choose to fix $\bm\pi_\a$ at the boundary, so that physical states are naturally described by wave functions of the boundary value of $\bm\pi_\a$.

Because the action \eqref{twistor action} is topological---with no Hamiltonian---the path integral reduces to an ordinary integral. Parametrizing the worldline $I$ by $\tau$ and discretizing, one finds
\ba
    && \int D\bm\xi\, D\bm\pi\,
    \la\Psi_2|\bm\pi(\tau_f)\ra
    \la\bm\pi(\tau_i)|\Psi_1\ra\,
    \exp\left(-i\int_{\tau_i}^{\tau_f}\dd\tau\,\partial_\tau \bm\pi_\a\cdot \bm\xi^\a \right) \nn
    &&=\lim_{N\to \infty}
    \int \frac{\dd^4 \bm\pi(\tau_0)}{(2\pi)^2}
    \left(\prod_{n=1}^N
    \frac{\dd^4 \bm\xi(\tau_n)}{(2\pi)^2}
    \frac{\dd^4\bm\pi(\tau_n)}{(2\pi)^2}\right)
     \la\Psi_2|\bm\pi(\tau_N)\ra  \la\bm\pi(\tau_0)|\Psi_1\ra
      \times \nn
      &&\hspace{60pt} \times\,
     \exp\left(-i\sum_{n=1}^N
    \left(\bm\pi_\a(\tau_n)-\bm\pi_\a(\tau_{n-1})\right)\cdot \bm\xi^\a(\tau_n)\right),
    \label{WLP reduction}
\ea
where the endpoints ${\rm s}I$ and ${\rm e}I$ correspond to $\tau_0$ and $\tau_N$, respectively. The $\bm\xi(\tau_n)$ integrals produce delta functions $\delta^4(\bm\pi_\a(\tau_n)-\bm\pi_\a(\tau_{n-1}))$, which subsequently  identify $\bm\pi(\tau_n)$ with $\bm\pi(\tau_{n-1})$. Performing these integrals for $n=1,\ldots,N$, the discretized path integral collapses to the single integral
\be
    \int \frac{\dd^4\bm\pi(\tau_0)}{(2\pi)^2}\, \la\Psi_2|\bm\pi(\tau_0)\ra  \la\bm\pi(\tau_0)|\Psi_1\ra\,,
\ee
where the $\tau_0$ dependence is immaterial since $\bm\pi(\tau_0)$ is itself an integration variable. The path integral therefore amounts to inserting the completeness relation,
\begin{align}
    \mathbb{I}=\int \frac{\dd^4\bm\pi}{(2\pi)^2}\,|\bm\pi\ra\la \bm\pi|\,,
\end{align}
between $\la\Psi_2|$ and $|\Psi_1\ra$.

Consequently, upon quantization, the Hilbert space $\mathcal{H}$ of the worldline theory \eqref{twistor action} is realized as the space of momentum eigenfunctions $\Psi(\bm\pi)=\la \bm\pi|\Psi\ra=\la \pi_1,\pi_2|\Psi\ra$, on which the position variables $\xi_1^\a$ and $\xi_2^\a$ act as the differential operators $i\,\partial/\partial\pi_{1\a}$ and $i\,\partial/\partial\pi_{2\a}$, respectively. The spectrum of \eqref{twistor action} contains \emph{twice} the field content of Vasiliev's HSG, namely the massless fields of spin $s=0,1,2,\ldots$. For each $s\ge 1$, the two copies of the spin-$s$ irrep arise as the eigenspaces of the \emph{spin} operator
\be\label{spin generator}
    \cS=\frac{1}{2}\Big(\pi_{1\a}\,\frac{\partial}{\partial \pi_{2\a}}-
    \pi_{2\a}\,\frac{\partial}{\partial \pi_{1\a}}\Big),
\ee
with eigenvalues $\pm s$.\footnote{The spin operator \eqref{spin generator} corresponds to the $\cS$ appearing in the constraint \eqref{spin const} of the constrained massless spin-$s$ action. Our worldline model is the unconstrained version, obtained by removing this constraint.}
We emphasize that the $\cS=\pm s$ eigenspaces do \emph{not} correspond to the helicity-$\pm s$ irreps, but rather to two copies of the same spin-$s$ irrep (see Appendix \ref{sec:Howe} for further discussion). The $\cS=0$ eigenspace, on the other hand, contains the scalar and pseudo-scalar irreps, which are inequivalent but share the same mass squared.

To construct the states created by AdS$_4$ field operators in this Hilbert space, we consider the AdS$_4$ isometry algebra $\mathfrak{so}(2,3)\simeq\mathfrak{sp}(4,\mathbb{R})$, generated by
\be
P_{\a\b}=\bm\pi_\a\cdot\bm\pi_\b\,,
\qquad
K^{\a\b}=-\frac{\partial}{\partial \bm\pi_\a}\cdot\frac{\partial}{\partial \bm\pi_\b}\,,
\qquad
L_\a{}^\b=-i\left(\bm\pi_\a\cdot \frac{\partial}{\partial \bm\pi_\b}+\d_\a^\b\right),
\label{sp4}
\ee
where $P_{\a\b}$ generates transverse translations with respect to the boundary, $D=\tfrac{1}{2}\,L_\a{}^\a$ generates radial translations, $J_{\a\b}=L_{(\a}{}^\g\,\epsilon_{\b)\g}$ generates transverse Lorentz transformations, and $B_{\a\b}=\tfrac{1}{2}(P_{\a\b}-K_{\a\b})$ generates boosts that mix the transverse and radial directions.\footnote{Equivalently, viewed as generators of three-dimensional conformal symmetry, $P_{\a\b}$ and $K^{\a\b}$ generate translations and special conformal transformations, while $J_{\a\b}$ and $D$ generate Lorentz transformations and dilatations. See Appendix \ref{sec: spinor convention} for further conventions.}
It will also be convenient to introduce the parity maps $\mathsf{P}_\a$ and $\mathsf{S}$, defined by
\be
    \mathsf{P}_1\,\binom{\pi_{I1}}{\pi_{I2}}=\binom{-\pi_{I1}}{\pi_{I2}}\,,
    \qquad
    \mathsf{P}_2\,\binom{\pi_{I1}}{\pi_{I2}}=\binom{\pi_{I1}}{-\pi_{I2}}\,,
    \qquad
    \mathsf{S}\,\binom{\pi_{I1}}{\pi_{I2}}=\binom{\pi_{I2}}{\pi_{I1}}\,.
    \label{parity}
\ee
The maps $\mathsf{P}_1$ and $\mathsf{P}_2$ flip the $\mu=1$ and $\mu=2$ spatial directions, respectively, and are related by an element of the identity component of $Sp(4,\mathbb{R})$. Throughout this paper, we use spinor indices to denote three-dimensional vectors; see Appendix \ref{sec: spinor convention} for the vector--spinor dictionary.

Since an AdS$_4$ field operator $\cO(x,\rho)$ transforms covariantly under AdS isometry, its value at any point is determined by its value at a reference point $\mathsf{o}$ as
\be
	\mathcal{O}(x,\rho)= e^{\frac{i}{2}\,x^{\a\b}\,P_{\a\b}}\,e^{-i\,\rho\,D}
	\,\mathcal{O}(\mathsf{o})\,e^{i\,\rho\,D}\,e^{-\frac{i}{2}\,x^{\a\b}\,P_{\a\b}}\,,
	\label{Cov}
\ee
where $x^{\a\b}$ and $\rho$ are the Poincar\'e patch coordinates,
\be
	\dd s^2=\dd\rho^2 +e^{2\rho}\,\dd x^2=\frac{\dd z^2+\dd x^2}{z^2}\,,
\ee
with the AdS radius set to unity for simplicity. The asymptotic boundary lies at $z=0$, i.e., $\rho=+\infty$. We use $z=e^{-\rho}$ and $\rho$ interchangeably whenever convenient.

\subsection{Scalar field in AdS$_4$}\label{sec:scalar}

As a first example, we compute the wave function $\la \bm\pi |\,\Phi(x,z)\,|\Omega\ra$ created by the scalar field operator $\Phi(x,z)$. Assuming AdS$_4$-invariance of the vacuum and using \eqref{Cov}, we find
\be
		\la \bm\pi |\,\Phi(x,z)\,|\Omega\ra
=e^{\frac{i}{2}\,x^{\a\b}\,\bm\pi_\a\cdot\bm\pi_\b}\,
		z\,\la \sqrt{z}\,\bm\pi |\,\Phi(\mathsf{o})\,|\Omega\ra\,,
\ee
which reduces the problem to evaluating the wave function $\la \bm\pi |\,\Phi(\mathsf{o})\,|\Omega\ra$ at the origin $\mathsf{o}=(\rho=0, x^{\a\b}=0)$. Imposing invariance under the transverse Lorentz generator $J_{\a\b}$ yields the differential equation
\be
	0=\la \bm\pi|\,[J_{\a\b},\Phi(\mathsf{o})]\,|\Omega\ra
	=i\,\bm\pi_{(\a}\cdot\frac{\partial}{\partial \bm\pi^{\b)}}\,
	\la \bm\pi |\,\Phi(\mathsf{o})\,|\Omega\ra\,,
\ee
whose general solution is
\be
	\la \bm\pi|\,\Phi(\mathsf{o})\,|\Omega\ra=f(\la\pi_1\pi_2\ra)\,,
\ee
where we use the notation $\la \chi\,\psi\ra =\epsilon^{\a\b}\,\chi_\a\,\psi_\b=\chi_\a\,\psi^\a$. Imposing in addition invariance under the boost generator $B_{\a\b}$ gives
\begin{align}
	0=\la \bm\pi |\,[B_{\a\b},\Phi(\mathsf{o})]\,|\Omega\ra
	&=\frac{1}{2} \left(\bm\pi_\a\cdot\bm\pi_\b
	+\frac{\partial}{\partial \bm\pi^\a}\cdot\frac{\partial}{\partial \bm\pi^\b}
	\right)\la \bm\pi |\,\Phi(\mathsf{o})\,|\Omega\ra \nn
	&= \frac{1}{2}\,\bm\pi_\a\cdot \bm\pi_\b
	\left( \frac{d^2}{dy^2}+1\right) f(y)\,\Big|_{y=\la \pi_1\pi_2\ra}\,.
	\label{B on scalar}
\end{align}
The general solution involves two arbitrary constants $c_+$ and $c_-$,
\be
	\la \bm\pi |\,\Phi(x,z)\,|\Omega\ra=
	e^{\frac{i}{2}\,x^{\a\b}\,\bm\pi_\a\cdot\bm\pi_\b}\,z
	\left[c_+\,e^{+i\,z\,\la\pi_1\pi_2\ra}
	+c_-\,e^{-i\,z\,\la\pi_1\pi_2\ra}\right]. \label{free scalar sol}
\ee
One can verify that \eqref{free scalar sol} satisfies the Klein--Gordon equation
\be
	\left[(z\,\partial_z)^2-3\,z\,\partial_z+z^{2}\,\partial^2_x-m^2\right]
	\la \bm\pi |\,\Phi(x,z)\,|\Omega\ra=0
	\qquad \text{with}\qquad m^2=-2\,,
\ee
where the mass squared is that of a conformal scalar in four dimensions.  Note that the solution \eqref{free scalar sol} transforms as $c_+\leftrightarrow c_-$ under the parity map. Therefore, $c_+=c_-$ defines a genuine (parity-even) scalar, while $c_+=-c_-$ defines a pseudo-scalar (parity-odd). These two fields may be identified with the scalars of type-A and type-B HSG in AdS$_4$, respectively.

\paragraph{Boundary conditions.}
From \eqref{free scalar sol} we read off the boundary behavior near $z=0$. The scalar and pseudo-scalar identifications correspond, respectively, to the Dirichlet and Neumann boundary conditions, with fall-offs $\mathcal{O}(z)$ and $\mathcal{O}(z^2)$:
\bs
 \begin{align}
	 {\rm Dirichlet}\ (c_+=+c_-):\quad
	\la \bm\pi |\,\Phi(x,z)\,|\Omega\ra& =
	2\,c_+\,e^{\frac{i}{2}\,x^{\a\b}\,\bm\pi_\a\cdot\bm\pi_\b}\,
	z\,\cosh(i\,z\,\la\pi_1\pi_2\ra)\nn
	&\simeq 2\,c_+\,z\,e^{\frac{i}{2}\,x^{\a\b}\,\bm\pi_\a\cdot\bm\pi_\b}\,,
\end{align}
\begin{align}
	{\rm Neumann}\ (c_+=-c_-):\quad
	\la \bm\pi |\,\Phi(x,z)\,|\Omega\ra&=
	2\,c_+\,e^{\frac{i}{2}\,x^{\a\b}\,\bm\pi_\a\cdot\bm\pi_\b}\,
	z\,\sinh(i\,z\,\la\pi_1\pi_2\ra)\nn
	&\simeq 2\,c_+\,z^2\,i\la \pi_1\pi_2\ra\,e^{\frac{i}{2}\,x^{\a\b}\,\bm\pi_\a\cdot\bm\pi_\b}\,,
\end{align}
\es
where $\simeq$ denotes the leading behavior as $z\to 0$.

\paragraph{Two-point functions.}
From the reality condition $\Phi(x,z)^\dagger=\Phi(x,z)$, the two-point function is
\begin{align}
	& \la \Omega|\,\Phi(x,z)\,\Phi(x',z')\,|\Omega\ra
	=\int\frac{\dd^4\bm\pi}{(2\pi)^2}\,
	\la \Omega|\,\Phi(x,z)\,|\bm\pi\ra\la\bm\pi|\,\Phi(x',z')\,|\Omega\ra \nn
	&=
	\int \frac{\dd^4\bm\pi}{(2\pi)^2}\,
	e^{-\frac{i}{2}\,(x^{\a\b}-x'^{\a\b})\,\bm\pi_\a\cdot\bm\pi_\b}\,
	z\,z'\, \Big( |c_+|^2\,e^{i\,(z'-z)\la\pi_1\pi_2\ra}+|c_-|^2\,e^{-i\,(z'-z)\la\pi_1\pi_2\ra}\nn
	&\hspace{165pt}
	+c_+^*\,c_-\,e^{-i\,(z'+z)\la\pi_1\pi_2\ra} +c_-^*\,c_+\,e^{i\,(z'+z)\la\pi_1\pi_2\ra}\Big)\,,
\end{align}
The Gaussian $\bm\pi$-integral gives
\be
	\la \Omega|\,\Phi(x,z)\,\Phi(x',z')\,|\Omega\ra
	=\frac{\frac{|c_+|^2+|c_-|^2}{2}}{{\rm Dist}^2(x,z|x',z')}+\frac{\frac{c_+^*\,c_-+c_-^*\,c_+}{2}}{{\rm Dist}^2(x,-z|x',z')}\,,
    \label{scalar Wightman}
\ee
where ${\rm Dist}^2(x,z|x',z')$ is the squared chordal distance between $(x,z)$ and $(x',z')$,
\be
	{\rm Dist}^2(x,z|x',z')=\frac{(x-x')^2+(z-z')^2}{2\,z\,z'}\,.
\ee
A noteworthy feature is the appearance of ${\rm Dist}^2(x,-z|x',z')$, the squared distance between $(x',z')$ and the point $(x,-z)$. The point $(x,-z)$ may be interpreted as the mirror image of $(x,z)$ across the asymptotic boundary, or equivalently as its antipodal image if global AdS is covered by two Poincar\'e charts with $z>0$ and $z<0$:
\be
	(x,z)_{\rm Mirror}=(x,-z)=(x,z)_{\rm Antipodal}\,.
\ee
Since AdS behaves as a box with a reflective boundary, both the Dirichlet and Neumann boundary conditions require the inclusion of this mirror (or antipodal) contribution \cite{Avis:1977yn,Allen:1985wd}; see \cite{Krotov:2005hq,Akhmedov:2020jsi} for a modern account of the related issues.

The boundary behavior of the two-point function is easily extracted. For the Dirichlet condition ($c_+=c_-$),
\bs
\be
	\la \Omega|\,\Phi(x,z)\,\Phi(x',z')\,|\Omega\ra
	\simeq \frac{|c_+|^2}{(x-x')^2}\,z\,z'\,,
\ee
while for the Neumann condition ($c_+=-c_-$),
\be
	\la \Omega|\,\Phi(x,z)\,\Phi(x',z')\,|\Omega\ra
	\simeq \frac{|c_+|^2}{(x-x')^4}\,z^2\,z'^2 \,.
\ee
\es

%%%%%%%%%%%%%%%%%%%%%%%%%%%%%%%%%%
\subsection{Spin-$s$ field in AdS$_4$}

We now turn to the wave functions created by massless spin-$s$ fields. Because of the gauge symmetries present for $s\ge 1$, the construction used above for the scalar does not uniquely determine the field, as explained in \cite{Weinberg:1995mt}. Nevertheless, we can still determine the corresponding Weyl tensors $C^{\sst (s)}_{\mu_1\ldots\mu_s,\nu_1\ldots \nu_s}$ of the totally symmetric rank-$s$ massless gauge fields $\varphi^{\sst (s)}_{\mu_1\ldots\mu_s}$ up to a coefficient ambiguity.

Recall that the Weyl tensors are gauge-invariant field strengths in the $(s,s)$ representation of of $\mathfrak{sl}(2,\mathbb{C})\simeq \mathfrak{so}(1,3)$, obtained by taking $s$ curls of a spin-$s$ gauge field and projecting onto the traceless part. In four dimensions, they decompose into self-dual (SD) and anti-self-dual (ASD) parts, carrying the $(2s,0)$ and $(0,2s)$ irreps, respectively. In spinor notation, these SD/ASD parts are symmetric rank-$2s$ tensors, with dotted and undotted indices related by Hermitian conjugation. By (anti-)self-duality, they are determined by their transverse components, which form symmetric rank-$2s$ tensors of $\mathfrak{sl}(2,\mathbb{R})\simeq \mathfrak{so}(1,2)$. Since the dotted/undotted distinction is then absent, we use undotted indices throughout, denoting the SD and ASD Weyl tensors by $C^{\sst (+s)}_{\a_1\cdots \a_{2s}}$ and $C^{\sst (-s)}_{\a_1\cdots \a_{2s}}$. Without loss of generality, we focus on the SD case, since
\be
	C^{\sst (-s)}_{\a_1\cdots \a_{2s}}=(C^{\sst (+s)}_{\a_1\cdots \a_{2s}})^\dagger\,.
	\label{C dagger}
\ee
Imposing self-duality alone would amount to setting the ASD fields to zero, leading to non-unitary results unless the signature is changed to $(0,4)$ or $(2,2)$.

As in the scalar case, AdS$_4$ covariance relates the Weyl tensor at any point to its value at the origin,
\be
	C^{\sst (+s)}_{\a_1\cdots \a_{2s}}(x,\rho)= e^{\frac{i}{2}\,x^{\a\b}\,P_{\a\b}}\,e^{-i\,\rho\,D}
	\,C^{\sst (+s)}_{\a_1\cdots \a_{2s}}(\mathsf{o})\,e^{i\,\rho\,D}\,e^{-\frac{i}{2}\,x^{\a\b}\,P_{\a\b}}\,,
\ee
and it suffices to determine $C^{\sst (+s)}_{\a_1\cdots \a_{2s}}(\mathsf{o})$ via bulk Lorentz covariance. The generators of $\mathfrak{sl}(2,\mathbb{C})$ are
\begin{equation}
	R_{\a\b}=J_{\a\b}+i\,B_{\a\b}\,,
	\qquad  \text{and} \qquad
	R_{\a\b}^\dagger\,,
\end{equation}
satisfying
\be
    [R_{\a\b},R_{\g\d}]=i\big(\epsilon_{\g(\a}\,R_{\b)\d}+\epsilon_{\d(\a}\,R_{\b)\g}\big)\,,\qquad [R_{\a\b},R_{\g\d}^\dagger]=0\,.
\ee
A SD spin-$s$ tensor, transforming in the $(2s,0)$ irrep, must obey
\bs
\begin{align}\label{eq:SD-condition-1}
	[R_{\a\b}, C^{\sst (+s)}_{\g_1\cdots \g_{2s}}(\mathsf{o})]&=
	\sum_{n=1}^{2s} \epsilon_{(\a|\g_n}\,C^{\sst (+s)}_{\g_1\cdots |\b)|\cdots \g_{2s}}(\mathsf{o})\,, \\
	[R_{\a\b}^\dagger,C^{\sst (+s)}_{\g_1\cdots \g_{2s}}(\mathsf{o})]&=0\,.
\end{align}
\es
Solving these conditions as in the scalar case gives
\begin{align}
	\la\bm\pi|\,C^{\sst (+s)}_{\a_1\cdots \a_{2s}}(\mathsf{o})\,|\Omega\ra
	&= c^{\sst (+s)}_{+}\,e^{+i\la\pi_1\pi_2\ra}\,
	(\pi_{1}+i\pi_{2})_{(\a_1}\cdots (\pi_{1}+i\pi_{2})_{\a_{2s})} \nn
	&\ \ \,+  c^{\sst (+s)}_{-}\,e^{-i\la\pi_1\pi_2\ra}
	(\pi_{1}-i\pi_{2})_{(\a_1}\cdots (\pi_{1}-i\pi_{2})_{\a_{2s})}\,,
\end{align}
with two undetermined coefficients $c^{\sst (+s)}_\pm$ for each spin. Since we focus on free fields in this section and different spins do not mix, we drop the spin dependence and write $c^{\sst (+s)}_\pm = c^{\sst (+)}_\pm$.

It is then convenient to introduce a generating function that packages all SD Weyl tensors together with the scalar,
\be
	C^{\sst (+)}(x,z;v)=\sum_{2s=0}^\infty\frac{1}{(2s)!}\,v^{\a_1}\cdots v^{\a_{2s}}\,C^{\sst (+s)}_{\a_1\cdots\a_{2s}}(x,z)\,,
\ee
from which a single spin-$s$ Weyl tensor may be extracted as
\begin{align}
    C^{\sst (+s)}_{\a_1\ldots\a_{2s}}(x,z)=\frac{\partial}{\partial v^{\a_{1}}}\ldots \frac{\partial}{\partial v^{\a_{2s}}}C^{\sst (+)}(x,z;v)\Big|_{v=0}\,.
\end{align}
Repeating the scalar-case treatment, the SD wave function reads
\bs
\begin{align}\label{wave function SD}
	&\la\bm\pi|\,C^{\sst (+)}(x,z;v)\,|\Omega\ra
	\nn
    &= e^{\frac{i}{2}\,x^{\a\b}\,\bm\pi_\a\cdot\bm\pi_\b}\,z\left[
	c^{\sst (+)}_{+}\,e^{+i\,z\la\pi_1\pi_2\ra+\sqrt{z}\la (\pi_1+i\pi_2) v\ra}
	+c^{\sst (+)}_{-}\,e^{-i\,z\la\pi_1\pi_2\ra+\sqrt{z}\la (\pi_1-i\pi_2) v\ra}\,
	\right],
\end{align}
and similarly for the ASD Weyl tensor,
\begin{align}\label{wave function ASD}
	&\la\bm\pi|\,C^{\sst (-)}(x,z;v)\,|\Omega\ra
	\nn
    &=e^{\frac{i}{2}\,x^{\a\b}\,\bm\pi_\a\cdot\bm\pi_\b}\,z\left[
    c^{\sst (-)}_{+}\,e^{+i\,z\la\pi_1\pi_2\ra+\sqrt{z}\la (\pi_1-i\pi_2) v\ra}
	+c^{\sst (-)}_{-}\,e^{-i\,z\la\pi_1\pi_2\ra+\sqrt{z}\la (\pi_1+i\pi_2) v\ra}\,
	\right].
\end{align}
\es
As in the scalar case, one verifies that the individual spin-$s$ wave functions satisfy the Bargmann--Wigner equations, the higher-spin generalization of the source-free Maxwell equations projected onto the SD and ASD sectors:
\bs
\begin{align}           
&\left[z\,\partial_{\a}{}^{\b_1}+i\left(z\partial_z+s+1\right)\d_{\a}^{\b_1} \right] \la \bm\pi|\,C_{\b_1\cdots \b_{2s}}^{\sst(+s)}(x,z)\,|\Omega\ra= 0\,,
    \label{Bargmann SD} \\
    &\left[z\,\partial_\a{}^{\b_1}-i\left(z\partial_z+s+1\right)\d_\a^{\b_1} \right] \la \bm\pi|\, C_{\b_1\cdots \b_{2s}}^{\sst(-s)}(x,z)\,|\Omega\ra= 0\,.
    \label{Bargmann ASD}
\end{align}
\es
Together with \eqref{eq:SD-condition-1}, this confirms that \eqref{wave function SD} and \eqref{wave function ASD} are the wave functions created by the SD and ASD spin-$s$ Weyl tensors.

\paragraph{Boundary conditions.}

Every field in AdS$_4$ must be supplemented by a suitable boundary condition. For massless spin-$s$ fields, the \emph{only} boundary condition consistent with AdS$_4$ covariance and unitarity is the Dirichlet condition, which sets the leading boundary behavior of the gauge potential to zero. A short proof of this statement will appear elsewhere; see also \cite{Marolf:2006nd,Compere:2008us} for related discussions in the gravitational case.

Working with Weyl tensors rather than gauge potentials, we must translate this condition appropriately. Near the boundary, the SD and ASD Weyl tensors behave as
\bs
\begin{align}
	&\la\bm\pi|\,C^{\sst (+)}(x,z;v)\,|\Omega\ra\simeq e^{\frac{i}{2}\,x^{\a\b}\,\bm\pi_\a\cdot\bm\pi_\b}
	\sum_{2s=0}^\infty
	z^{s+1}\left(c^{\sst (+)}_+\,\tfrac{\la (\pi_1+i\pi_2) v\ra^{2s}}{(2s)!}
	+c^{\sst (+)}_-\,\tfrac{\la (\pi_1-i\pi_2) v\ra^{2s}}{(2s)!}\right),\\
   & \la\bm\pi|\, C^{\sst (-)}(x,z;v)\,|\Omega\ra
    \simeq e^{\frac{i}{2}\,x^{\a\b}\,\bm\pi_\a\cdot\bm\pi_\b}
	\sum_{2s=0}^\infty
	z^{s+1}\left(c^{\sst (-)}_+\,\tfrac{\la (\pi_1-i\pi_2) v\ra^{2s}}{(2s)!}
	+c^{\sst (-)}_-\,\tfrac{\la (\pi_1+i\pi_2) v\ra^{2s}}{(2s)!}\right).
\end{align}
\es
The ``electric'' and ``magnetic'' parts of the Weyl tensor,
\be\label{E and M def}
    C^{\sst (\rm E)}=\tfrac{1}{2}\left( C^{\sst (+)}+C^{\sst (-)}\right),\qquad
    C^{\sst (\rm M)}=\tfrac{i}{2}\left(C^{\sst (+)}-C^{\sst (-)}\right),
\ee
exhibit the same fall-off $\mathcal{O}(z^{s+1})$ but correspond to different boundary objects. The boundary value of $C^{\sst (\rm E)}_{\a_1\cdots \a_{2s}}$ encodes the physical degrees of freedom and corresponds, via the holographic dictionary, to a traceless conserved current $J^{\a_1\cdots\a_{2s}}$ in CFT$_3$. The boundary value of $C^{\sst (\rm M)}_{\a_1\cdots \a_{2s}}$, on the other hand, is the higher-spin generalization of the Cotton tensor \cite{Pope:1989vj, Henneaux_2016} of the boundary value of the gauge potential $\varphi^{\sst (s)}_{\a_1\cdots \a_{2s}}$, which sources the current $J^{\a_1\cdots \a_{2s}}$.

Imposing the Dirichlet boundary condition amounts to setting the boundary value of the magnetic Weyl tensor to zero, which relates the coefficients as
\be
    c^{\sst (+)}_+=c^{\sst (-)}_-=c_+\,,
    \qquad
    c^{\sst (+)}_-=c^{\sst (-)}_+=c_-\,.
\ee
Note that two undetermined constants $c_+$ and $c_-$ remain even after imposing the boundary condition. This ambiguity reflects the fact that the Hilbert space of our worldline model contains the spin-$s$ irrep twice.

\paragraph{Two-point functions.}

We now compute the two-point functions. The mixed-helicity two-point function reads
\ba
	&&\la\Omega|\,C^{\sst (-)}(x,z;v)\,C^{\sst (+)}(x',z';v')\,|\Omega\ra=\nn
	&&=
	\dfrac{(|c_+|^2+|c_-|^2)\, e^{-2i\,\sqrt{zz'}\, \la  v|P(x,z;x',z')| v'\ra}}
	{2\,{\rm Dist}^2(x,z;x',z')}
	+\dfrac{c_+^*c_-+c_-^*c_+}
	{2\,{\rm Dist}^2(x,-z;x',z')}\,,
	\label{-+ 2pt}
\ea
where the last term is the scalar contribution, and the $P_{\a\b}$ structure is
\be\label{2-point P}
	P_{\a\b}(x,z;x',z')=\frac{(x-x')_{\a\b}+i\,(z-z')\,\epsilon_{\a\b}}{
	(x-x')^2+(z-z')^2}\,,
\ee
where we used the notation
$\la v|\,V\,|w\ra=v^\a\,V_{\a\b}\,w^\b$.
This two-point function describes the propagation of positive-helicity states created by the SD fields. 
The negative-helicity propagation
can be computed in a similar manner:
\ba
    && \la\Omega|\,C^{\sst (+)}(x,z;v)\,C^{\sst (-)}(x',z';v')\,|\Omega\ra \nn 
    &&=
	\dfrac{(|c_+|^2+|c_-|^2)\, e^{-2i\,\sqrt{zz'}\, \la  v|\bar P(x,z;x',z')| v'\ra}}
	{2\,{\rm Dist}^2(x,z;x',z')}
	+\dfrac{c_+^*c_-+c_-^*c_+}
	{2\,{\rm Dist}^2(x,-z;x',z')}\,.
	\label{+- 2pt} 
\ea
Interestingly, the same-helicity two-point functions,
\ba
	&&\la\Omega|\,C^{\sst (+)}(x,z;v)\,C^{\sst (+)}(x',z';v')\,|\Omega\ra=\nn
	&&=
	\dfrac{(|c_+|^2+|c_-|^2)\, e^{-2i\,\sqrt{zz'}\, \la  v|\bar P(x,-z;x',z')| v'\ra}}
	{2\,{\rm Dist}^2(x,-z;x',z')}
	+\dfrac{c_+^*c_-+c_-^*c_+}
	{2\,{\rm Dist}^2(x,z;x',z')}\,,
    \label{++ 2pt}
\ea
and its complex conjugate $\la \Omega|\,C^{\sst (-)}\,C^{\sst (-)}\,|\Omega\ra$, do not vanish. As is evident from their structure, these non-vanishing same-helicity correlators may be interpreted as mirror-image contributions of the mixed-helicity ones: a massless spin-$s$ field can flip its helicity upon reflection off the boundary.

The boundary behavior of these two-point functions deserves careful analysis. To this end, it is useful to work with the rescaled Weyl tensor $\tfrac{1}{z}\,C^{\sst (\pm)}(x,z;\tfrac{v}{\sqrt{z}})$, which makes the boundary limit more transparent. The rescaling introduces a factor of $1/z^{s+1}$ for each Weyl tensor, yielding the leading behavior
\begin{align}
&\lim_{z,z'\to 0}\la\Omega|\,\tfrac{1}{z}\,C^{\sst (-)}(x,z;\tfrac{v}{\sqrt{z}})\,\tfrac{1}{z'}\,C^{\sst (+)}(x',z';\tfrac{v'}{\sqrt{z'}})\,|\Omega\ra\nn
 &=
	(|c_+|^2+|c_-|^2)
\left[
\frac{\exp\left(-2i\,\la v|\tfrac{x-x'}{(x-x')^2}|v'\ra\right) }{(x-x')^2}+
2\,\pi^2\,\la v\,v'\ra\,F\left(\la v|\tfrac{\partial}{\partial x} |v'\ra\right)\,
    \delta^3(x-x')\right],
\end{align}
with 
\be 
    F(y)=\sum_{n=0}^\infty 
    \frac{i^n\,y^n}{n!\,(n+1)!}
    =\frac{I_1(2\sqrt{i\,y})}{\sqrt{i\,y}}\,.
\ee
The second term is a local contact term arising from the boundary limit,
\begin{align}
    \lim_{z\to 0}\frac{x_{\a_1\b_1}\cdots x_{\a_{n}\b_{n}}\,z}{(x^2+z^2)^{2+n}}
    =\frac{(-2)^{-n}\,\pi^2}{(n+1)!}\,\partial_{\a_1\b_1}\cdots \partial_{\a_{n}\b_{n}}\,\delta^3(x)\,.
\end{align}
This contact term is parity-odd and can be interpreted as a higher-spin extension of Chern--Simons theory: for spin $s$, it arises from the contraction of the spin-$s$ gauge potential with the corresponding $(2s-1)$-derivative higher-spin Cotton tensor on the boundary \cite{Pope:1989vj,Henneaux_2016}. This boundary local functional may itself emerge from a higher-spin generalization of a topological bulk action of the type $\int F\wedge F$ in four dimensions.

Ultimately, we are interested in the correlation functions of the electric Weyl tensor $C^{\sst\rm (E)}$, whose boundary limits correspond to the conserved current correlators in CFT$_3$. Combining the various contributions, the two-point function
\be
    \la \Omega|\,C^{\sst (\rm E)}\,C^{\sst (\rm E)}\,|\Omega\ra= \frac{1}{4}
\sum_{\s=\pm}\sum_{\s'=\pm}\la \Omega|\,C^{\sst (\s)}\,C^{\sst (\s')}\,|\Omega\ra
\ee
has boundary limit
\be
\lim_{z,z'\to 0}\la\Omega|\,\tfrac{1}{z}\,C^{\sst (\rm E)}(x,z;\tfrac{v}{\sqrt{z}})\,\tfrac{1}{z'}\,C^{\sst (\rm E)}(x',z';\tfrac{v'}{\sqrt{z'}})\,|\Omega\ra
 =
	\frac{|c_+|^2+|c_-|^2}{(x-x')^2}\,\exp\left[-2i\,\la v|\tfrac{x-x'}{(x-x')^2}|v'\ra\right].
\ee
The parity-odd contact terms cancel, in agreement with the parity invariance of the theory.

%%%%%%%%%%%%%%%%%%%%%%%%%%%%%%%%%%%%%%%%%%%%%%%%%%%%%%
\section{Worldline model as interacting theory}
\label{sec:3}

Having treated the free worldline model in Section \ref{sec:2}, we now turn to the question of how interactions can be introduced in a string-inspired manner. In string theory, interactions are geometrically encoded in the joining and splitting of worldsheets. Adopting an analogous picture at the level of worldlines provides a natural and intuitive framework for introducing interaction vertices into our model.

Let us consider a situation involving three states rather than two. The transition from the joining of two incoming states $|\Psi_1\ra$ and $|\Psi_3\ra$ to a single outgoing state $\la\Psi_2|$ should be given by the path integral
\begin{align}\label{test process}
 	A= \int D\bm Z\, \la\Psi_2|\bm\pi({\rm e}I_2)\ra\,\la\bm\pi({\rm s}I_1)|\Psi_1\ra\,
    \la\bm\pi({\rm s}I_3)|\Psi_3\ra\,
    \exp\left(i\int_{I_1\cup I_2 \cup I_3}
    \O_{AB}\,\bm Z^A\cdot \dd \bm Z^B\right),
\end{align}
with appropriate boundary terms understood. Diagrammatically, this transition can be depicted as
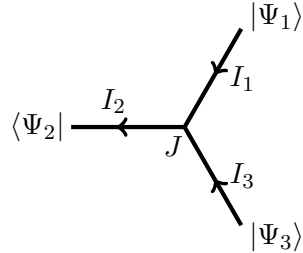
\begin{center}
\begin{tikzpicture}[thick]
\coordinate (O) at (0,0);
\draw[ultra thick] (60:0) -- (60:0.9);\draw[ultra thick,<-] (60:0.8) -- (60:1.5);
\draw[ultra thick] (180:1.5) -- (180:0.8);\draw[ultra thick,<-] (180:0.9) -- (180:0);
\draw[ultra thick,->] (300:1.5) -- (300:0.8);\draw[ultra thick] (300:0.9) -- (300:0);
\node at (40:1) {$I_1$};
\node at (160:1) {$I_2$};
\node at (320:1) {$I_3$};
\node at (50:1.9) {$|\Psi_1\ra$};
\node at (310:1.9) {$|\Psi_3\ra$};
\node at (180:1.95) {$\la \Psi_2|$};
\node at (-0.15,-0.25) {$J$};
\end{tikzpicture}
\end{center}
which contains the junction $J=I_1\cap I_2\cap I_3$. For the path integral to be well-defined, suitable junction conditions must be imposed on the worldline fields $\bm Z^A$ at
\be
    J={\rm e}I_1={\rm s}I_2={\rm e}I_3\,.
\ee
Because the worldline action is first-order in derivatives, only half  of the boundary values $\bm Z^A({\rm e}I_1)$, $\bm Z^A({\rm s}I_2)$, and $\bm Z^A({\rm e}I_3)$ should be related at the junction.

%%%%%%%%%%%%%%%%%%%%%%%%%%%%%%%%%%%%%%%%%%
\subsection{Junction condition}

String theory itself is free at the level of first quantization, with interactions arising from the joining conditions imposed on the worldsheet fields at junctions. By analogy, in the HS worldline theory we therefore seek a prescription that smoothly glues three worldline fields at the junction. Our key observation is that the worldline variables $\bm Z^A$ are in fact ``two-twistors,'' $\bm Z^A=(Z^A_1,Z_2^A)$, and each ``one-twistor'' $Z_i^A$ enjoys all the covariant properties of the AdS$_4$ symmetry. Therefore, $Z_1^A$ and $Z_2^A$ can be freely separated as
\ba
	\int D\bm Z\ e^{i\int_{I_i}
    \O_{AB}\,\bm Z^A\cdot \dd \bm Z^B}
    \eq \int DZ_1\,DZ_2\
    e^{i
    \int_{I_{i}}
    \O_{AB}\,Z^A_1 \,\dd Z^B_1
    +i\int_{I_{i}}
    \O_{AB}\,Z^A_2 \,\dd Z^B_2}\nn
     \eq \int DZ\
    e^{i
    \int_{\color{blue!58!black!100}I_{i}^{\sst \rm B}}
    \O_{AB}\,Z^A \,\dd Z^B
    +i\int_{\color{red!58!black!100}I_{i}^{\sst \rm R}}
    \O_{AB}\,Z^A \,\dd Z^B}\,,
\ea
where the worldline $\color{blue!58!black!100}I_{i}^{\sst \rm B}$ is associated with $Z^A_1$ and $\color{red!58!black!100}I_{i}^{\sst \rm R}$ with $Z_2^A$; both are duplicates of the original interval $I_i$. The advantage of this prescription is that we can work with single-twistor fields $Z^A({\color{blue!58!black!100}I^{\sst \rm B}}/
{\color{red!58!black!100}I^{\sst \rm R}})$ instead of two-twistor fields $\bm Z^A(I)$, at the cost of working on a double-line diagram:
\begin{center}
\begin{tikzpicture}[thick]
\draw[ultra thick] (0,0)--(1.5,0);\draw[ultra thick,<-] (1.4,0)--(3,0);
\node at (1.5,0.35) {$\bm Z^A(I)$};
\node at (4.5,0) {$\Longrightarrow$};
\begin{scope}[shift={(6,0)}]
\draw[blue!58!black!100] (0,0.12)--(1.5,0.12);\draw[blue!58!black!100,<-] (1.45,0.12)--(3,0.12);
\draw[red!58!black!100] (0,-0.12)--(1.5,-0.12);\draw[red!58!black!100,<-] (1.45,-0.12)--(3,-0.12);
\node at (1.5,0.5) {$Z^A({\color{blue!58!black!100} I^{\sst B}})$};
\node at (1.5,-0.5) {$Z^A({\color{red!58!black!100} I^{\sst R}})$};
\end{scope}
\end{tikzpicture}
\end{center}
Returning to the transition amplitude \eqref{test process}, the double-line description allows the following junction conditions:
\be
        {\rm e}{\color{blue!58!black!100}I_{1}^{\sst\rm B}}={\rm s}{\color{blue!58!black!100}I_{2}^{\sst \rm B}}\,,
        \qquad
        {\rm  s}{\color{red!58!black!100}I_{2}^{\sst \rm R}}={\rm  e}{\color{red!58!black!100}I_{3}^{\sst \rm R}}\,,
        \qquad
        {\rm e}{\color{blue!58!black!100}I_{3}^{\sst \rm B}}={\rm e}{\color{red!58!black!100}I_{1}^{\sst \rm R}}\,.
\ee
Since the endpoints of $\color{blue!58!black!100}I_{1}^{\sst \rm B}$ and $\color{red!58!black!100}I_{3}^{\sst \rm R}$ match with the starting points of $\color{blue!58!black!100}I_{2}^{\sst \rm B}$ and $\color{red!58!black!100}I_{2}^{\sst \rm R}$, respectively, they form the full arcs $\color{blue!58!black!100}I_{\sst \vv{12}}^{\sst \rm B}$ and $\color{red!58!black!100}I_{\sst \vv{32}}^{\sst \rm R}$. Here, the arrows over the indices indicate the flow of the worldline fields:
\begin{center}
\begin{tikzpicture}[thick]
\draw[blue,shift={(150:0.1)}] (60:0.05) -- (60:.9);\draw[blue!58!black!100,<-,shift={(150:0.1)}] (60:0.8) -- (60:1.5);
\draw[red!58!black!100,shift={(150:-0.1)}] (60:0.05) -- (60:.9);\draw[red!58!black!100,<-,shift={(150:-0.1)}] (60:0.8) -- (60:1.5);
\draw[blue!58!black!100,shift={(270:-0.1)}] (180:1.5) -- (180:.9);\draw[blue!58!black!100,shift={(270:-0.1)},<-] (180:1) -- (180:.05);
\draw[red!58!black!100,shift={(270:0.1)}] (180:1.5) -- (180:.9);\draw[red!58!black!100,shift={(270:0.1)},<-] (180:1) -- (180:.05);
\draw[blue!58!black!100,shift={(-150:-0.1)}] (-60:0.05) -- (-60:.9);\draw[blue!58!black!100,<-,shift={(-150:-0.1)}] (-60:0.8) -- (-60:1.5);
\draw[red!58!black!100,shift={(-150:0.1)}] (-60:0.05) -- (-60:.9);\draw[red!58!black!100,<-,shift={(-150:0.1)}] (-60:0.8) -- (-60:1.5);
\node at (30:1) {\color{red!58!black!100}$I_{1}^{\sst \rm R}$};\node at (90:1) {\color{blue!58!black!100}$I_{1}^{\sst \rm B}$};
\node at (155:1) {\color{blue!58!black!100}$I_{2}^{\sst \rm B}$};\node at (205:1) {\color{red!58!black!100}$I_{2}^{\sst \rm R}$};
\node at (-30:1) {\color{blue!58!black!100}$I_{3}^{\sst \rm B}$};\node at (-90:1) {\color{red!58!black!100}$I_{3}^{\sst \rm R}$};
\node at (2.5,0) {$\Longrightarrow$};
\begin{scope}[shift={(5,0)}]
\draw[blue!58!black!100,<-,shift={(150:0.1)}] (60:0.8) -- (60:1.5);
\draw[blue!58!black!100,shift={(270:-0.1)}] (180:0.95) arc (-90:-30:1.55);
\draw[blue!58!black!100,shift={(270:-0.1)},-<] (180:1.5) -- (180:.9);
\draw[red!58!black!100,shift={(270:0.1)},-<] (180:1.5) -- (180:.9);
\draw[red!58!black!100,shift={(270:0.1)}] (180:0.95) arc (90:30:1.55);
\draw[red!58!black!100,<-,shift={(-150:0.1)}] (-60:0.8) -- (-60:1.5);
\draw[red!58!black!100,<-,shift={(150:-0.1)}] (60:0.8) -- (60:1.5);
\draw[red!58!black!100,shift={(150:-0.1)}] (60:0.95) arc (150:180:1.55);
\draw[blue!58!black!100,shift={(-150:-0.1)}] (-60:0.95) arc (210:180:1.55);
\draw[blue!58!black!100,<-,shift={(-150:-0.1)}] (-60:0.8) -- (-60:1.5);
\node at (120:0.7) {\color{blue!58!black!100}$I_{\sst \vv{12}}^{\sst \rm B}$};
\node at (240:0.7) {\color{red!58!black!100}$I_{\sst \vv{32}}^{\sst \rm R}$};
\node at (30:1) {\color{red!58!black!100}$I_{1}^{\sst \rm R}$};
\node at (-30:1) {\color{blue!58!black!100}$I_{3}^{\sst \rm B}$};
\end{scope}
\end{tikzpicture}
\end{center}
Notice that the worldlines $\color{red!58!black!100}I_{1}^{\sst \rm R}$ and $\color{blue!58!black!100}I_{3}^{\sst \rm B}$ have opposite orientations, so they cannot be glued unless one of them is reversed. Choosing to flip $\color{blue!58!black!100}I_{3}^{\sst \rm B}$,
\begin{center}
\begin{tikzpicture}[thick]
\draw[blue!58!black!100,->] (0,0)--(1,0);\draw[blue!58!black!100] (1,0)--(2,0);
\node at (1,0.35) {\color{blue!58!black!100}$I_{3}^{\sst \rm B}$};
\node at (3,0) {$\Longrightarrow$};
\begin{scope}[shift={(4,0)}]
\draw[red!58!black!100] (0,0)--(1,0);\draw[red!58!black!100,<-] (1,0)--(2,0);
\node at (1,0.35) {\color{red!58!black!100}$I_{3}^{\sst \rm R}$};
\end{scope}
\end{tikzpicture}
\end{center}
we assume that the color also changes upon reversal of orientation. At the level of the worldline action, the orientation flip introduces an overall minus sign in the one-twistor action,
\be
    \int_{\color{blue!58!black!100}I_{3}^{\sst \rm B}}\Omega_{AB}\,Z^A\,\dd Z^B
    =- \int_{\color{red!58!black!100}I_{3}^{\sst \rm R}}\Omega_{AB}\,Z^A\,\dd Z^B\,.
\ee
To glue $\color{red!58!black!100}I_{1}^{\sst \rm R}$ and $\color{red!58!black!100}I_{3}^{\sst \rm R}$ into a connected line $\color{red!58!black!100}I_{\sst\vv{13}}^{\sst \rm R}$, this minus sign must be absorbed by a redefinition of $Z^A$ on $\color{red!58!black!100}I_{3}^{\sst \rm R}$. We make the minimal choice,
\begin{align}
    Z^A\quad \to \quad i\,Z^A\,.
\end{align}
Under this redefinition, all spacetime isometry generators flip sign,
\be
    P_{\a\b}, L_{\a}{}^\b, K^{\a\b}
    \quad \to \quad
     -P_{\a\b}, -L_{\a}{}^\b, -K^{\a\b}\,,
\ee
as expected, since reversing the arrow of a worldline corresponds to backward propagation of the worldline particle. Applying these rules, the original transition amplitude becomes
\ba
    A &=& \int D\xi^\a D\pi_\a\
    \la\Psi_2|\pi({\rm e}{\color{blue!58!black!100}I_{\sst\vv{12}}^{\sst \rm B}}),
    \pi({\rm e}{\color{red!58!black!100}I_{\sst \vv{32}}^{\sst \rm R}})\ra\,
    \la\pi({\rm s}{\color{blue!58!black!100}I_{\sst \vv{12}}^{\sst \rm B}}),\pi({\rm s}{\color{red!58!black!100}I_{\sst \vv{13}}^{\sst \rm R}})|\Psi_1\ra \times \nn
    && \hspace{50pt} \times\,
    \la i\,\pi({\rm e}{\color{red!58!black!100}I_{\sst \vv{13}}^{\sst \rm R}}),\pi({\rm s}{\color{red!58!black!100}I_{\sst\vv{32}}^{\sst \rm R}})|\Psi_3\ra\,
    \exp\left(-i\int_{L^{\sst \rm {\color{blue!58!black!100}B}{\color{red!58!black!100}RR}}} \dd\pi_\a\, \xi^\a\right),
\ea
over the worldline $L^{\sst \rm {\color{blue!58!black!100}B}{\color{red!58!black!100}RR}}$, which can be depicted as
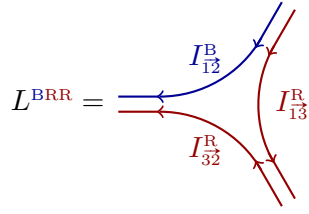
\begin{center}
\begin{tikzpicture}[thick]
\node at (-2.3,0.06) {$L^{\sst \rm {\color{blue!58!black!100}B}{\color{red!58!black!100}RR}}=$};
\draw[blue!58!black!100,<-,shift={(150:0.1)}] (60:0.8) -- (60:1.5);
\draw[blue!58!black!100,shift={(270:-0.1)}] (180:0.95) arc (-90:-30:1.55);
\draw[blue!58!black!100,shift={(270:-0.1)},-<] (180:1.5) -- (180:.9);
\draw[red!58!black!100,shift={(270:0.1)},-<] (180:1.5) -- (180:.9);
\draw[red!58!black!100,shift={(270:0.1)}] (180:0.95) arc (90:30:1.55);
\draw[red!58!black!100,<-,shift={(-150:0.1)}] (-60:0.8) -- (-60:1.5);
\draw[red!58!black!100,<-,shift={(150:-0.1)}] (60:0.8) -- (60:1.5);
\draw[red!58!black!100,->,shift={(150:-0.1)}] (60:0.95) arc (150:210:1.55);
\draw[red!58!black!100,shift={(-150:-0.1)}] (-60:0.8) -- (-60:1.5);
\node at (120:0.7) {\color{blue!58!black!100}$I_{\sst \vv{12}}^{\sst \rm B}$};
\node at (240:0.7) {\color{red!58!black!100}$I_{\sst \vv{32}}^{\sst \rm R}$};
\node at (0:0.8) {\color{red!58!black!100}$I_{\sst \vv{13}}^{\sst {\rm R}}$};
\end{tikzpicture}
\end{center}
Repeated application of the above rules allows us to reexpress the same transition amplitude as
\ba
     A&=&\int D\xi^\a D\pi_\a\
    \la \Psi_2|\,i\,\pi({\rm s}{\color{red!58!black!100}I_{\sst\vv{21}}^{\sst \rm R}}),\pi({\rm e}{\color{red!58!black!100}I_{\sst \vv{32}}^{\sst \rm R}})\ra\,
    \la i\,\pi({\rm e}{\color{red!58!black!100}I_{\sst \vv{21}}^{\sst \rm R}}),\pi({\rm s}{\color{red!58!black!100}I_{\sst \vv{13}}^{\sst \rm R}})\,|\Psi_1\ra\times\nn
    && \hspace{50pt} \times\,
    \la i\,\pi({\rm e}{\color{red!58!black!100}I_{\sst \vv{13}}^{\sst \rm R}}),
    \pi({\rm s}{\color{red!58!black!100}I_{\sst\vv{32}}^{\sst \rm R}})\,|\Psi_3\ra\,
    \exp\left(-i\int_{L^{\sst \rm {\color{red!58!black!100}RRR}}} \dd\pi_\a\,\xi^\a\right)
    \nn
     &=&  \int D\xi^\a D\pi_\a\
    \la \Psi_2|\,\pi({\rm e}{\color{blue!58!black!100}I_{\sst\vv{12}}^{\sst \rm B}}),i\,\pi({\rm s}{\color{blue!58!black!100}I_{\sst \vv{23}}^{\sst \rm B}})\ra\,
    \la \pi({\rm s}{\color{blue!58!black!100}I_{\sst \vv{12}}^{\sst \rm B}}),i\,\pi({\rm e}{\color{blue!58!black!100}I_{\sst \vv{31}}^{\sst \rm B}})\,|\Psi_1\ra\times\nn
    && \hspace{50pt} \times\,
    \la\pi({\rm s}{\color{blue!58!black!100}I_{\sst \vv{31}}^{\sst \rm B}}),
    i\,\pi({\rm e}{\color{blue!58!black!100}I_{\sst\vv{23}}^{\sst \rm B}})\,|\Psi_3\ra\,
    \exp\left(-i\int_{L^{\sst \rm {\color{blue!58!black!100}BBB}}} \dd\pi_\a\,\xi^\a\right),
\ea
either over $L^{\sst \rm {\color{red!58!black!100}RRR}}$ (red lines only) or over $L^{\sst \rm {\color{blue!58!black!100}BBB}}$ (blue lines only):
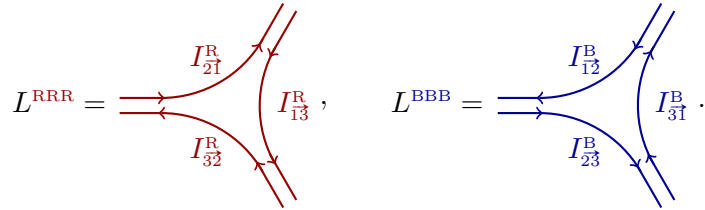
\begin{center}
\begin{tikzpicture}[thick]
\node at (-2.3,0.06) {$L^{\sst \rm {\color{red!58!black!100}RRR}}=$};
\draw[red!58!black!100,>-,shift={(150:0.1)}] (60:0.8) -- (60:1.5);
\draw[red!58!black!100,shift={(270:-0.1)}] (180:0.95) arc (-90:-30:1.55);
\draw[red!58!black!100,shift={(270:-0.1)},->] (180:1.5) -- (180:.9);
\draw[red!58!black!100,shift={(270:0.1)},-<] (180:1.5) -- (180:.9);
\draw[red!58!black!100,shift={(270:0.1)}] (180:0.95) arc (90:30:1.55);
\draw[red!58!black!100,<-,shift={(-150:0.1)}] (-60:0.8) -- (-60:1.5);
\draw[red!58!black!100,<-,shift={(150:-0.1)}] (60:0.8) -- (60:1.5);
\draw[red!58!black!100,->,shift={(150:-0.1)}] (60:0.95) arc (150:210:1.55);
\draw[red!58!black!100,shift={(-150:-0.1)}] (-60:0.8) -- (-60:1.5);
\node at (120:0.7) {\color{red!58!black!100}$I_{\sst \vv{21}}^{\sst \rm R}$};
\node at (240:0.7) {\color{red!58!black!100}$I_{\sst \vv{32}}^{\sst \rm R}$};
\node at (0:0.8) {\color{red!58!black!100}$I_{\sst \vv{13}}^{\sst \rm R}$};
\node at (1.2,0) {$,$};
\begin{scope}[shift={(5,0)}]
\node at (-2.3,0.06) {$L^{\sst \rm {\color{blue!58!black!100}BBB}}=$};
\draw[blue!58!black!100,<-,shift={(150:0.1)}] (60:0.8) -- (60:1.5);
\draw[blue!58!black!100,shift={(270:-0.1)}] (180:0.95) arc (-90:-30:1.55);
\draw[blue!58!black!100,shift={(270:-0.1)},-<] (180:1.5) -- (180:.9);
\draw[blue!58!black!100,shift={(270:0.1)},->] (180:1.5) -- (180:.9);
\draw[blue!58!black!100,shift={(270:0.1)}] (180:0.95) arc (90:30:1.55);
\draw[blue!58!black!100,>-,shift={(-150:0.1)}] (-60:0.8) -- (-60:1.5);
\draw[blue!58!black!100,>-,shift={(150:-0.1)}] (60:0.8) -- (60:1.5);
\draw[blue!58!black!100,shift={(-150:-0.1)}] (-60:0.95) arc (210:150:1.55);
\draw[blue!58!black!100,<-,shift={(-150:-0.1)}] (-60:0.8) -- (-60:1.5);
\node at (120:0.7) {\color{blue!58!black!100}$I_{\sst \vv{12}}^{\sst \rm B}$};
\node at (240:0.7) {\color{blue!58!black!100}$I_{\sst \vv{23}}^{\sst \rm B}$};
\node at (0:0.8) {\color{blue!58!black!100}$I_{\sst \vv{31}}^{\sst \rm B}$};
\node at (1.2,0) {$.$};
\end{scope}
\end{tikzpicture}
\end{center}
In this way, we obtain a cyclic diagram with three insertion points. Choosing the blue-line convention, we find
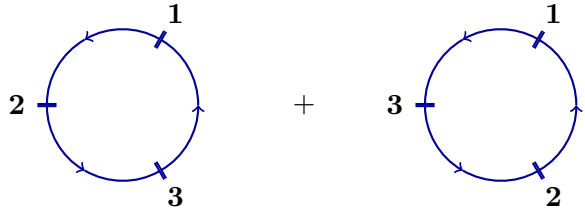
\begin{center}
\begin{tikzpicture}[thick]
\begin{scope}[shift={(2.5,0)}]
\draw[blue!58!black!100,|->] (60:1) arc (60:120:1);\draw[blue!58!black!100,-|] (120:1) arc (120:180:1);
\draw[blue!58!black!100,|->] (180:1) arc (180:240:1);\draw[blue!58!black!100,-|] (240:1) arc (240:300:1);
\draw[blue!58!black!100,|->] (300:1) arc (300:360:1);\draw[blue!58!black!100,-|] (0:1) arc (0:60:1);
\node at (60:1.4) {$\bm 1$};
\node at (180:1.4) {$\bm 3$};
\node at (300:1.4) {$\bm 2$};
\end{scope}
\node at (-0.1,0) {$+$};
\begin{scope}[shift={(-2.5,0)}]
\draw[blue!58!black!100,|->] (60:1) arc (60:120:1);\draw[blue!58!black!100,-|] (120:1) arc (120:180:1);
\draw[blue!58!black!100,|->] (180:1) arc (180:240:1);\draw[blue!58!black!100,-|] (240:1) arc (240:300:1);
\draw[blue!58!black!100,|->] (300:1) arc (300:360:1);\draw[blue!58!black!100,-|] (0:1) arc (0:60:1);
\node at (60:1.4) {$\bm 1$};
\node at (180:1.4) {$\bm 2$};
\node at (300:1.4) {$\bm 3$};
\end{scope}
\end{tikzpicture}
\end{center}
The first diagram corresponds to the construction just carried out, while the second represents the other inequivalent way of connecting the three worldlines. Since the two diagrams are equally valid, both contributions are to be summed.

\subsection{Vertex operators}\label{sec: vertex operator}

So far we have considered the transition amplitude between $\la \Psi_2|$ and $|\Psi_1\ra\otimes |\Psi_3\ra$ for a three-point interaction. We could equally well consider the transition between $\la \Psi_2|\otimes \la\Psi_3|$ and $|\Psi_1\ra$, and the two should be related by crossing symmetry. In string theory, crossing symmetry is manifest by virtue of the worldsheet description, in which a local vertex operator creates each string state. In this section, we explore whether an analogous vertex operator can be defined in our worldline theory.

Recall the picture developed in the previous section. Starting from a transition amplitude involving two states connected by a double line,
\begin{center}
\begin{tikzpicture}[thick]
\draw[blue!58!black!100] (-0.8,-0.4) arc (-90:-158:1.2);
\draw[blue!58!black!100,<-] (-0.8,-0.4) arc (-90:-35:1.2) ;
\draw[red!58!black!100] (-0.8,-0.1) arc (-90:-150:0.9);
\draw[red!58!black!100,<-] (-0.8,-0.1) arc (-90:-35:0.9);
\node at (-2.1,0.6) {$ \la\Psi_2| \pi_1,\pi_2\ra$};
\draw[red!58!black!100,<-] (1,0.9) arc (90:23:1.2);
\draw[red!58!black!100] (1,0.9) arc (90:145:1.2) ;
\draw[blue!58!black!100,<-] (1,0.6) arc (90:30:0.9);
\draw[blue!58!black!100] (1,0.6) arc (90:145:0.9);
\node at (2.3,-0.05) {$ \la \pi_1,\pi_2|\Psi_1\ra$};
\end{tikzpicture}
\end{center}
we apply the map that reverses the flow of one of the worldlines, taking red to blue:
\begin{center}
\begin{tikzpicture}[thick]
\draw[blue!58!black!100] (-0.8,-0.4) arc (-90:-158:1.2);
\draw[blue!58!black!100,<-] (-0.8,-0.4) arc (-90:-35:1.2) ;
\draw[blue!58!black!100,<-] (-0.8,-0.1) arc (-90:-150:0.9);
\draw[blue!58!black!100] (-0.8,-0.1) arc (-90:-35:0.9);
\node at (-2.1,0.6) {$ \la\Psi_2| \pi_1,i\pi_2\ra$};
\draw[blue!58!black!100] (1,0.9) arc (90:23:1.2);
\draw[blue!58!black!100,<-] (1,0.9) arc (90:145:1.2) ;
\draw[blue!58!black!100,<-] (1,0.6) arc (90:30:0.9);
\draw[blue!58!black!100] (1,0.6) arc (90:145:0.9);
\node at (2.3,-0.05) {$ \la \pi_1,i\pi_2|\Psi_1\ra$};
\end{tikzpicture}
\end{center}
which introduces a factor of $i$ in $\pi_2$. Now that a worldline enters and exits from the same state $\Psi$, we can interpret this as the action of an operator on that worldline. Accordingly, we define the vertex operator corresponding to a state $\Psi$ by
\be
    \la \pi_1|\,\hat V_\Psi\,|\pi_2\ra
    \equiv \la \pi_1,i\pi_2|\Psi\ra\,,
    \label{vertex}
\ee
which may be regarded as the worldline counterpart of the state--operator correspondence in two-dimensional CFT:
\begin{equation}\label{state-operator}
\begin{tikzpicture}[thick]
\begin{scope}[shift={(-3,0)}]
\draw[blue!58!black!100, |->] (-0.13,0) -- (-.13,0.6);\draw[blue!58!black!100] (-.13,0.6) -- (-.13,1.1);
\draw[red!58!black!100, |->] (.13,0) -- (.13,0.6);\draw[red!58!black!100] (.13,0.6) -- (.13,1.1);
\node at (0.25,-0.2) {$\la \pi_1,\pi_2|\Psi\ra$};
\end{scope}
\node at (-1,0.5) {$\Longleftrightarrow$};
\begin{scope}[shift={(1,0)}]
\draw[blue!58!black!100, |->] (-0.13,0) -- (-.13,0.6);\draw[blue!58!black!100] (-.13,0.6) -- (-.13,1.1);
\draw[blue!58!black!100,|-] (.13,0) -- (.13,0.5);\draw[blue!58!black!100,<-] (.13,0.5) -- (.13,1.1);
\node at (0.3,-0.2) {$\la \pi_1,i\pi_2|\Psi\ra$};
\end{scope}
\node at (3,0.5) {$\Longleftrightarrow$};
\begin{scope}[shift={(5.5,0)}]
\draw[blue!58!black!100, |->] (0,0.5) -- (-0.6,0.5);\draw[blue!58!black!100] (-0.6,0.5) -- (-1.1,0.5);
\draw[blue!58!black!100,|-] (0,0.5) -- (.5,0.5);\draw[blue!58!black!100,<-] (.5,0.5) -- (1.1,0.5);
\node at (0,0) {$\la \pi_1|\hat V_{\Psi}|\pi_2\ra$};
\end{scope}
\end{tikzpicture}
\end{equation}
Note that while the state $|\Psi\ra$ carries a unitary representation, the vertex operator $\hat V_\Psi$ carries an adjoint one. A symmetry generator $X$ accordingly acts as
\be
	X\rhd |\Psi\ra=X\,|\Psi\ra\,,
	\qquad
	X\rhd \hat V_{\Psi}=[X,\hat V_{\Psi}]\,,
\ee
reproducing the relation between the twisted-adjoint and adjoint representations of HSG, similar to \cite{Engquist:2005yt}.

From the definition \eqref{vertex}, the transition amplitude can be expressed as a trace,
\be
	\la \Psi_2|\Psi_1\ra=
	-\tr\left(\hat\tau\,\hat V^\dagger_{\Psi_2}\,\hat V_{\Psi_1}^{\phantom\dagger}\right),\label{tau map}
\ee
where $\hat\tau^\dagger=\hat\tau$ acts as $\hat \tau\,|\pi\ra=|-\pi\ra$, and the trace is defined by
\be
    \tr\,\hat\cO=\int \frac{\dd^2\pi'}{2\pi}\,\la\pi'|\,\hat\cO\,|\pi'\ra\,.
\ee
The expression \eqref{tau map} uses the identity $\la \Psi|\pi_1,i\pi_2\ra =\la \pi_2|\,\hat\tau\,\hat V^\dagger_\Psi\,|\pi_1\ra$. Note that the $\hat\tau$-inserted trace lacks manifest cyclic symmetry, whereas the simple trace
\be
    \tr\left(\hat V_{\Psi_2}\,\hat V_{\Psi_1}\right)
\ee
is manifestly cyclic. Before assessing the viability of this simple trace as a candidate ``scattering amplitude'' of HSG, we must first resolve an issue. Namely, for a given state $|\Psi\ra$, there can be two associated vertex operators, $\hat V_\Psi$ and $\hat V_\Psi^\dagger$. Unless these two operators are proportional to each other, 
an ambiguity remains.

To examine the Hermitian properties of the vertex operators corresponding to the higher-spin states constructed in the previous section, we begin with the state created by the real scalar, $|\Psi\ra=\Phi(x,z)\,|\Omega\ra$. The corresponding vertex operator is
\ba
    &&\la\pi_1|\,\hat V(x,z)\,|\pi_2\ra
    \equiv \la \pi_1,i\pi_2|\,\Phi(x,z)\,|\Omega\ra\nn
   &&= e^{\frac{i}{2}\,x^{\a\b}\,(\pi_{1\a}\pi_{1\b}-\pi_{2\a}\pi_{2\b})}\,z
	\left[c_+\,e^{-z\,\la\pi_1\pi_2\ra}
	+c_-\,e^{+z\,\la\pi_1\pi_2\ra}\right],
\ea
and its Hermitian conjugate is
\begin{align}
	\la\pi_1|\,\hat V(x,z)^\dagger\,|\pi_2\ra
	&=\la\pi_2|\,\hat V(x,z)\,|\pi_1\ra^*\nn
    &=e^{\frac{i}{2}\,x^{\a\b}\,(\pi_{1\a}\pi_{1\b}-\pi_{2\a}\pi_{2\b})}\,z
	\left[c_+^*\,e^{+z\,\la\pi_1\pi_2\ra}
	+c_-^*\,e^{-z\,\la\pi_1\pi_2\ra}\right].
\end{align}
Requiring the proportionality
\be
    \hat V(x,z)^\dagger=w\,\hat V(x,z)\,,
\ee
we find
\be
    w=\frac{c_+^*}{c_-}=\frac{c_-^*}{c_+}\qquad
   \Rightarrow \qquad |c_+|=|c_-|\,.
\ee
This is consistent with the two boundary conditions: the scalar with the Dirichlet condition corresponds to $c_+=c_-$, and the pseudo-scalar with the Neumann condition to $c_+=-c_-$.

The trace of two vertex operators can be computed by Gaussian integration, with the result
\be \label{two-point vertex}
	\tr\left(\hat V(x,z)\,\hat V(x',z') \right)
	=\pm c_+^2\left(\frac{1}{{\rm Dist}^2(x,z|x',z')}\pm\frac{1}{{\rm Dist}^2(x,-z|x',z')}\right),
\ee
where the $\pm$ signs correspond to the Dirichlet and Neumann conditions, respectively.
Comparison with \eqref{scalar Wightman} yields the two choices
\ba
	&\text{Scalar with Dirichlet BC:}\qquad &c_+=c_-=c\,,\nn
        &\text{Pseudo-scalar with Neumann BC:}\qquad &c_+=-c_-=i\,c\,,
\ea
with real $c$, rendering the vertex operator Hermitian with $w=1$.

Turning to the higher-spin states, we define the corresponding vertex operators as
\bs
\begin{align}
	&
     \la\pi_1|\hat V^{\sst (+)}(x,z;v)|\pi_2\ra
    =
    \la\pi_1,i\pi_2|\,C^{\sst (+)}(x,z;v)\,|\Omega\ra
	\nn
    &= e^{\frac{i}{2}\,x^{\a\b}
    (\pi_{1\a}\pi_{1\b}-\pi_{2\a}\pi_{2\b})}\,z\left[
	c^{\sst (+)}_{+}\,e^{-\,z\la\pi_1\pi_2\ra+\sqrt{z}\la (\pi_1-\pi_2) v\ra}
	+c^{\sst (+)}_{-}\,e^{z\la\pi_1\pi_2\ra+\sqrt{z}\la (\pi_1+\pi_2) v\ra}\,
	\right],\\
	&\la\pi_1|\hat V^{\sst (-)}(x,z;v)|\pi_2\ra=
    \la\pi_1,i\pi_2|\,C^{\sst (-)}(x,z;v)\,|\Omega\ra
	\nn
    &=e^{\frac{i}{2}\,x^{\a\b}\,(\pi_{1\a}\pi_{1\b}
    -\pi_{2\a}\pi_{2\b})}\,z\left[
    c^{\sst (-)}_{+}\,e^{-z\la\pi_1\pi_2\ra+\sqrt{z}\la (\pi_1+\pi_2) v\ra}
	+c^{\sst (-)}_{-}\,e^{z\la\pi_1\pi_2\ra+\sqrt{z}\la (\pi_1-\pi_2) v\ra}\,
	\right].
\end{align}
\es
Requiring
\be
	\hat V^{\sst (+)}(x,z;v)^\dagger=w\,\hat V^{\sst (-)}(x,z;v)
\ee
gives, for integer spin (i.e., for even powers of $v$),
\be
	w=\frac{c^{\sst (+)}_+{}^*}{c^{\sst (-)}_-}
	=\frac{c^{\sst (+)}_-{}^*}{c^{\sst (-)}_+}\,,
\ee
which is again consistent with the magnetic boundary condition $c^{\sst (+)}_+=c^{\sst (-)}_-=c_+$ and $c^{\sst (+)}_-=c^{\sst (-)}_+=c_-$, yielding
\be
	w=\frac{c_+^*}{c_+}=\frac{c_-^*}{c_-}\,.
\ee
The traces of two vertex operators are
\bs
\ba
	&&\tr\left(\hat V^{\sst (\pm)}(x,z;v)\,\hat V^{\sst (\pm)}(x',z';v')\right)=\dfrac{c_+^2+c_-^2}
	{{\rm Dist}^2(x,z;x',z')}+ \nn 
    &&\hspace{60pt} +\,c_+\,c_-\,
	\dfrac{e^{\mp 2i\,\sqrt{zz'}\, \la  v| P(x,-z;x',z')| v'\ra}
    +e^{\pm 2i\,\sqrt{zz'}\, \la  v|\bar P(x,-z;x',z')| v'\ra
    }}{\,{\rm Dist}^2(x,-z;x',z')}\,,	
    \label{2pt tr ++}
\ea 
\ba
	&& \tr\left(\hat V^{\sst (\pm)}(x,z;v)\,\hat V^{\sst (\mp)}(x',z';v')\right)=\dfrac{c_+^2+c_-^2}
	{{\rm Dist}^2(x,-z;x',z')}+\nn 
    &&\hspace{60pt}+
    \,c_+\,c_- \,
	\dfrac{e^{\mp 2i\,\sqrt{zz'}\, \la  v|P(x,z;x',z')| v'\ra}
    +e^{\pm 2i\,\sqrt{zz'}\, \la  v|\bar P(x,z;x',z')| v'\ra}}{
	{\rm Dist}^2(x,z;x',z')}\,.
    \label{2pt tr +-}
\ea
\es
Once again, 
comparing with \eqref{-+ 2pt}, \eqref{+- 2pt}
and \eqref{++ 2pt},\footnote{Although the trace formulas 
\eqref{2pt tr ++} and \eqref{2pt tr +-}
generate the two-point functions for all spins, the relative coefficients $c_\pm$ are in fact independent for each spin, so the comparison with  \eqref{-+ 2pt}, \eqref{+- 2pt}
and \eqref{++ 2pt} must be carried out spin by spin.}
we are left with two classes of admissible vertex operators,
up to an overall real factor $c$, 
\be\label{c+ c-}
	(c_+,c_-,w)=\left(c\,e^\l\,,
    c\,e^{-\l},1\right) \quad \text{or} \quad
	\left(i\,c\,e^{\l}\,,-i\,c\,e^{-\l},-1\right),
\ee
with a residual one-parameter freedom within each class, parametrized by an arbitrary $\lambda \in \mathbb{R}$.

Finally, we note a subtle issue: the Gaussian integral appearing in the trace formula is not convergent in any analytic region of the time coordinates, in contrast to the two-point transition amplitude considered in the previous section. The latter is the Wightman function, whose integral representation involves $e^{i\,(t-t')\,\o}$ with $\o=\tfrac{1}{2}\,\delta^{\a\b}(\pi_{1\a}\pi_{1\b}+\pi_{2\a}\pi_{2\b})\ge 0$, so that the integral converges for ${\rm Im}(t-t')>0$. For the trace of vertex operators, by contrast, the corresponding frequency $\o=\tfrac{1}{2}\,\delta^{\a\b}(\pi_{1\a}\pi_{1\b}-\pi_{2\a}\pi_{2\b})$ has no definite sign, and no analytic continuation in $t-t'$ suffices to render the integral convergent. This pattern is suggestive of---though not literally identical to---the structure of the Feynman propagator, and a cleaner interpretation appears to require passing to Euclidean signature, with the twistor variables analytically continued in tandem. We do not pursue this analysis here, but hope to return to it in future work.

\paragraph{Operator form.}

So far, we have identified the matrix elements of the vertex operators that close into a worldline loop. It is also instructive to derive their symbols and operator expressions. The symbol of a matrix element is obtained via the Wigner--Weyl transform,
\be
    V(x,z|\xi,\pi)
    =\int \frac{\dd^2\varpi}{2\pi}\,e^{-i\,\varpi_\a\,\xi^\a}
    \left\la \pi+\tfrac{\varpi}{2}\right|
    \hat V(x,z)\left|\pi-\tfrac{\varpi}{2}\right\ra.
\ee
For the scalar field, one finds 
\be
    V(x,z|\xi,\pi)
    =c_+\,z\,\delta^2(\xi^\a-x^{\a\b}\,\pi_\b
    -iz\,\pi^\a)
    +c_-\,z\,
    \delta^2(\xi^\a-x^{\a\b}\,\pi_\b
    +iz\,\pi^\a)\,.
\ee
The delta functions impose the relations $\xi^{\a}=x^{\a\b}\,\pi_{\b}\pm iz\,\pi^{\a}$, which can be recognized as twistor incidence relations.\footnote{A somewhat similar structure, $\xi^\a=z^{-1/2}\,K\,\big[(x-\mathsf{x})^{\a}{}_{\b}\,v^{\b}\mp iz\,v^{\a}\big]$, with $\mathsf{x}^{\a\b}$ a boundary reference point and $K$ the bulk-to-boundary propagator of a scalar field of dimension $\Delta=1$, was employed in \cite{Didenko:2012tv}. We comment further on this connection below.} The appearance of the imaginary unit suggests that an analytic continuation of the real twistor space is required. The corresponding operator $\hat V(x,z)$ is obtained by promoting the symbol in the standard way:
\be
    \hat V(x,z)
    =c_+\,z\,\delta^2(\hat \xi^\a-x^{\a\b}\,\hat\pi_\b
    -iz\,\hat \pi^\a)
    +c_-\,z\,
    \delta^2(\hat \xi^\a-x^{\a\b}\,\hat\pi_\b
    +iz\,\hat \pi^\a)\,,
\ee
where the delta function with operator argument is understood as
\be
    \delta^2(\hat{\cO})
    =\int \frac{\dd^2\varpi}{2\pi}\,e^{i\,\la \varpi\,\hat\cO\ra}\,.
\ee
For the spinning case, the symbols read
\ba
     V^{\sst {(\pm)}}(x,z|\xi,\pi)
   &=& c_+\,z\,\d^{2}(\xi^\a -x^{\a\b}\,\pi_\b\pm i\,z\,\pi^\a+i \sqrt{z}\,v^\a) \nn
   &&+\,c_-\,z\,\d^{2}(\xi^\a -x^{\a\b}\,\pi_\b\mp i\,z\,\pi^\a)\,e^{2\sqrt{z}\,\la \pi\,v\ra}\,,
\ea
with the corresponding operator forms
\ba
    \hat V^{\sst {(\pm)}}(x,z)
   &=& c_+\,z\,\d^{2}(\hat \xi^\a -x^{\a\b}\,\hat \pi_\b\,
   \pm i\,z\,\hat \pi^\a+i \sqrt{z}\,v^\a) \nn
   &&+\,c_-\,z\,\d^{2}(\hat \xi^\a -x^{\a\b}\,\hat \pi_\b \mp i\,z\,\hat \pi^\a+i \sqrt{z}\,v^\a
   )\,e^{2\sqrt{z}\,\la \hat \pi\,v\ra}\,.
\ea
Note that upon passing from the symbol to the operator, the delta-function argument is shifted by $i\,\sqrt{z}\,v^\a$, so that the incidence relation itself acquires a polarization-dependent shift.

\subsection{Superselection}

We have seen that, for massless spin-$s$ fields, the corresponding vertex operator is not uniquely fixed, even after imposing the Dirichlet boundary condition---which amounts to setting the magnetic field to zero---together with the (anti-)Hermiticity, and even after accounting for the overall real factor. The remaining ambiguity is parametrized by $e^{2\lambda}$, the ratio between $c_+$ and $c_-$ for each spin, where $c_\pm$ are the coefficients of the $\cS=\pm s$ eigenspaces. This ambiguity, present at every spin, originates from the fact that the Hilbert space of our worldline model contains the HSG spectrum twice---once for each sign of the $\cS$ eigenvalue. Were these ambiguities left unresolved, the theory would suffer from an infinite tower of free parameters, none of which carry physical meaning.

We therefore eliminate them by employing the dual parity maps
\bs
\be
    \tilde{\mathsf{P}}_1(\pi_{1\a},\pi_{2\a})=(-\pi_{1\a},\pi_{2\a})\,,
	\qquad
	\tilde{\mathsf{P}}_2(\pi_{1\a},\pi_{2\a})=(\pi_{1\a},-\pi_{2\a})\,,
    \label{tilde P}
\ee
\be
	\tilde{\mathsf{S}}\,(\pi_{1\a},\pi_{2\a})=(\pi_{2\a},\pi_{1\a})\,.
    \label{tilde S}
\ee
\es
Unlike the physical parity maps \eqref{parity}, these $\mathbb{Z}_2$ maps preserve every AdS$_4$ isometry generator but reverse the sign of the spin operator:
\be
    \tilde{\mathsf{P}}_I\,\cS\,\tilde{\mathsf{P}}_I^{-1}
    =\tilde{\mathsf{S}}\,\cS\,\tilde{\mathsf{S}}^{-1}
    =-\cS\,.
\ee
The maps $\tilde{\mathsf{P}}_I$ and $\tilde{\mathsf{S}}$ are all equivalent up to conjugation by an $SO(2)$ rotation generated by $\cS$, so any one of them may be used to take a quotient of our Hilbert space. This quotient identifies the $\cS=+s$ and $\cS=-s$ eigenspaces, thereby resolving the ambiguity. Being $\mathbb{Z}_2$, the dual parity map admits only two distinct actions: trivial or sign-flipping. These two options correspond to type-A and type-B HSG, respectively. 
At the level of the vertex operator, we can use the $\tilde{\mathsf{P}}_2$ map to implement the $\mathbb{Z}_2$ quotient, so that the two cases read
\bs
\ba
    &\text{Type A:}
    \qquad &
    \la \pi_1|\,\hat V\,|-\pi_2\ra=+
    \la \pi_1|\,\hat V\,|\pi_2\ra\,,\label{A Oper}\\
    &\text{Type B:}
    \qquad &
    \la \pi_1|\,\hat V\,|-\pi_2\ra=-
    \la \pi_1|\,\hat V\,|\pi_2\ra\,. \label{B Oper}
\ea
\es
This sets $\lambda=0$ for every spin $s>0$ and uniquely fixes, up to the overall factor $c$, the vertex-operator coefficients---for all spins, including the (pseudo-)scalar---as
\bs
\ba
    &\text{Type A:}
    \qquad &
   c_+=+c_-=c\,,\label{A coefficients}\\
    &\text{Type B:}
    \qquad &
  c_+=-c_-=i\,c\,. \label{B coefficients}
\ea
\es
The preceding discussion was framed in terms of the worldline model \eqref{twistor action} on a doubled line: the target twistor space $\mathbb{T}$ of AdS$_4$, with coordinates $(Z_1{}^A, Z_2{}^A)$, is to be replaced by the orbifold $\mathbb{T}/\mathbb{Z}_2$ defined by the identification $(Z_1{}^A, Z_2{}^A)\sim (Z_1{}^A, -Z_2{}^A)$. Since the doubled line will eventually be split into two single lines, this orbifold construction must also be extended to the single-line case---a point to which we return in Section \ref{sec:5}.

\section{Worldline model for higher-spin gravity}\label{sec:4}

Building on the preceding discussion, we now propose a worldline formulation of HSG. Just as string theory furnishes S-matrix elements as its physical observables, the HS worldline theory furnishes correlation functions as its physical observables. The connected correlation functions of the quantum fields of HSG are given by
\ba\label{correlation functions def}
	&& \big\la\, C(x_1,z_1;v_1)\,\cdots\, C(x_n,z_n;v_n)\,\big\ra^{\rm HSG}_{\text{connected}}\nn
	&&
	=\sum_{\s\in S_n/\mathbb{Z}_n}
	\frac{1}{g^2_{\rm HSG}}\,
    \big\la\,\hat V(x_{\s(1)},z_{\s(1)};v_{\s(1)})\,\cdots\,\hat V(x_{\s(n)},z_{\s(n)};v_{\s(n)})\,\big\ra_{\rm WL}\,,
    \label{HSG correlator}
\ea
where the fields $C$ are either the scalar $\Phi$ or the electric Weyl tensor $C^{\sst\rm (E)}$, and $\hat V$ denotes the corresponding vertex operator. We have introduced the HSG coupling constant $g^{\phantom{g}}_{\rm HSG}$ as a proportionality constant relating the correlators of higher-spin fields to those of the vertex operators.\footnote{Since HSG should contain an Einstein gravity sector in some form, one expects $g_{\rm HSG}^2\propto G_{\rm N}\,L_{\rm AdS}^{-2}$, where $G_{\rm N}$ is the Newton constant and $L_{\rm AdS}$ is the AdS radius.} 

The summation runs over all distinct ways of connecting the $2n$ oriented worldlines---originating from the $n$ doubled-line insertions via the state--operator correspondence \eqref{state-operator}---into a single connected diagram representing the closed worldline loop $L$,
\begin{equation}
 \begin{tikzpicture}[thick,baseline,vertex/.style={anchor=base,
            circle,fill=black!25,minimum size=18pt,inner sep=2pt}]
            \draw[blue!58!black!100,|-] (2:1.2) arc (270:250:1.8);
            \draw[blue!58!black!100,|-] (-2:1.2) arc (90:110:1.8);
            \draw[blue!58!black!100,|-] (62:1.2) arc (330:310:1.8);
            \draw[blue!58!black!100,|-] (58:1.2) arc (150:170:1.8);
            \draw[blue!58!black!100,|-] (122:1.2) arc (30:10:1.8);
            \draw[blue!58!black!100,|-] (118:1.2) arc (210:230:1.8);
            \draw[blue!58!black!100,|-] (182:1.2) arc (90:70:1.8);
            \draw[blue!58!black!100,|-] (178:1.2) arc (270:290:1.8);
            \draw[blue!58!black!100,|-] (242:1.2) arc (150:130:1.8);
            \draw[blue!58!black!100,|-] (238:1.2) arc (330:350:1.8);
            \draw[blue!58!black!100,dotted] (270:1.2) arc (270:330:1.2);
             \filldraw[black!22!white!90!] (0,0) circle (0.6);
            \node at (0:1.5) {$\st{n}$};
            \node at (60:1.45) {$\st{1}$};
            \node at (120:1.45) {$\st{2}$};
            \node at (180:1.5) {$\st 3$};
            \node at (240:1.45) {$\st 4$};
        \end{tikzpicture}
    \qquad \Longleftrightarrow \qquad
    \sum_{\s\in S_n/\mathbb{Z}_n}
        \begin{tikzpicture}[thick,baseline,vertex/.style={anchor=base,
            circle,fill=black!25,minimum size=18pt,inner sep=2pt}]
            \draw[blue!58!black!100,|->] (0:1) arc (0:30:1);\draw[blue!58!black!100,-|] (30:1) arc (30:60:1);
            \draw[blue!58!black!100,|->] (60:1) arc (60:90:1);\draw[blue!58!black!100,-|] (90:1) arc (90:120:1);
            \draw[blue!58!black!100,|->] (120:1) arc (120:150:1);\draw[blue!58!black!100,-|] (150:1) arc (150:180:1);
            \draw[blue!58!black!100,|->] (180:1) arc (180:210:1);\draw[blue!58!black!100,-|] (210:1) arc (210:240:1);
            \draw[blue!58!black!100,dotted,|-|] (240:1) arc (240:360:1);
            \node at (0:1.5) {$\st{\s(n)}$};
            \node at (60:1.45) {$\st{\s(1)}$};
            \node at (120:1.45) {$\st \s(2)$};
            \node at (180:1.5) {$\st \s(3)$};
            \node at (240:1.45) {$\st \s(4)$};
            \node at (300:1.25) {$\color{blue!58!black!100} L$};
        \end{tikzpicture},
        \label{WL rule}
\end{equation}
where $S_n$ and $\mathbb{Z}_n$ are the symmetric and cyclic groups, respectively. 
The correlator of vertex operators is then given by the path integral
over the worldline $L$,
\ba
    &&
	\Big\la\,\hat V(x_{1},z_{1};v_{1})\,\cdots\,\hat V(x_{n},z_{n};v_{n})\,\Big\ra_{\rm WL}\nn
    &&
    =\int D\xi^\a\,D\pi_\a\,
    V(x_{1},z_{1};v_{1})\,\cdots\, V(x_{n},z_{n};v_{n})\,
    e^{-i\int_{L}\dd\pi_\a\,\xi^\a}
    \nn
    &&=\int \frac{\dd^2 \pi_{1}\cdots \dd^2\pi_n}{(2\pi)^n}\,
    \la \pi_{n}|\,\hat V(x_{1},z_{1};v_{1})\,|\pi_1\ra
   \cdots
    \la \pi_{n-1}|\,\hat V(x_{n},z_{n};v_{n})\,|\pi_{n}\ra\nn
    && =\tr\left(\hat V(x_{1},z_{1};v_{1})\,\cdots\,\hat V(x_{n},z_{n};v_{n})\right),
    \label{start off Gaussian}
\ea
which, as in the two-point case
\eqref{WLP reduction}, reduces to an ordinary finite-dimensional integral and coincides with the trace of the vertex operators.

\subsection{Bulk correlation functions}

The type-A and type-B vertex operators are defined by \eqref{A Oper} and \eqref{B Oper}, with the coefficients given by \eqref{A coefficients} and \eqref{B coefficients}. Since we restrict our attention to the integer-spin sector, only even powers of the auxiliary polarization $v$ contribute, and we replace each exponential $e^{\sqrt{z}\la (\pi_1\pm\pi_2)v\ra}$ by its even projection $\cosh(\sqrt{z}\la(\pi_1\pm\pi_2)v\ra)$. With the standard normalization of two-point functions, we choose the overall factor as $c=g^{\phantom g}_{\rm HSG}/2$. For the type-A theory, this yields
\bs
\ba
	 && \la\pi_1|\,\hat V_{\rm A}(x,z;v)\,|\pi_2\ra
     \nn
    &&= g^{\phantom g}_{\rm HSG}\,z\, e^{\frac{i}{2}\,x^{\a\b}
    (\pi_{1\a}\pi_{1\b}-\pi_{2\a}\pi_{2\b})}\,
    \cosh(z\la\pi_1\pi_2\ra)\,
    \cosh(\sqrt{z}\la \pi_1 v\ra) \,
    \cosh (\sqrt{z}\la \pi_2 v\ra)\,,
      \label{V type-A}
\ea
which also encompasses the scalar sector, since the two $\cosh$ factors involving $v$ reduce to unity at $v=0$. The corresponding vertex operator of the type-B theory reads 
\ba
	 && \la\pi_1|\,\hat V_{\rm B}(x,z;v)\,|\pi_2\ra\nn
    &&=i \,g^{\phantom g}_{\rm HSG}\,z\,e^{\frac{i}{2}\,x^{\a\b}
    (\pi_{1\a}\pi_{1\b}-\pi_{2\a}\pi_{2\b})}\,
    \cosh(z\la\pi_1\pi_2\ra)\,
    \sinh(\sqrt{z}\la \pi_1 v\ra) \,
    \sinh (\sqrt{z}\la \pi_2 v\ra)\,.
    \label{V type-B}
\ea
Unlike its type-A counterpart, this expression vanishes in the spin-zero sector. The pseudo-scalar field $\Phi(x,z)$ must therefore be accommodated by a separate vertex operator,
\be
	 \la\pi_1|\,\hat V_{\rm B}(x,z)\,|\pi_2\ra
	 = i\,g^{\phantom g}_{\rm HSG}\,z\,e^{\frac{i}{2}\,x^{\a\b}
    (\pi_{1\a}\pi_{1\b}-\pi_{2\a}\pi_{2\b})}\,
    \sinh(z\la\pi_1\pi_2\ra)\,,
    \label{V type-B 0}
\ee
\es
in which the overall sign factor has been redefined for later convenience.

To evaluate the $n$-point traces of these vertex operators, it is advantageous to first cast their matrix elements into a uniform Gaussian form. To this end, we introduce the auxiliary Gaussian vertex operator
\be
    \la\pi_1|\,\hat G(x,z;v^{\rm\sst L},v^{\rm \sst R})\,|\pi_2\ra
    =e^{\frac{i}{2}\,x^{\a\b}
    (\pi_{1\a}\pi_{1\b}-\pi_{2\a}\pi_{2\b})
   +z\la\pi_1\pi_2\ra +\la \pi_1 v^{\sst \rm L}\ra
   +\la \pi_2 v^{\sst \rm R}\ra}\,,
\ee
which serves as a generating object from which each physical vertex operator is recovered by an appropriate projection. Concretely, after rescaling $v\to v/\sqrt{z}$, one finds
\bs
\be
     \tfrac{1}{z}\,\hat V_{\rm A}\big(x,z;\tfrac{v}{\sqrt{z}}\big)
     =g^{\phantom g}_{\rm HSG}\,\hat G(x,z;v^{\sst\rm L},v^{\sst\rm R})\Big|_{\substack{\text{even}\,z\\
   \text{even}\,v_{\sst\rm L}, v_{\sst\rm R}\\
   v_{\sst\rm L}=v_{\sst\rm R}=v}}\,,
   \label{rescale A}
\ee
for the type-A vertex operator, and
\be
     \tfrac{1}{z}\,\hat V_{\rm B}\big(x,z;\tfrac{v}{\sqrt{z}}\big)
     =i\,g^{\phantom g}_{\rm HSG}\,\hat G(x,z;v^{\sst\rm L},v^{\sst\rm R})\Big|_{\substack{\text{even}\,z\\
   \text{odd}\,v_{\sst\rm L}, v_{\sst\rm R}\\
   v_{\sst\rm L}=v_{\sst\rm R}=v}}\,,
    \quad
     \tfrac{1}{z}\,\hat V_{\rm B}(x,z)
     =i\,g^{\phantom g}_{\rm HSG}\,\hat G(x,z;0,0)\Big|_{\text{odd}\,z}\,,
    \label{rescale B}
\ee
\es
for the spinning and pseudo-scalar sectors of the type-B theory, respectively. The key advantage of this reformulation is that the trace $\tr(\hat V\cdots\hat V)$ is obtained from $\tr(\hat G\cdots\hat G)$ by the same projection, and the latter is amenable to direct evaluation via matrix Gaussian integrals; see Appendix \ref{sec: Gaussian integral} for details.

The Gaussian trace assumes the standard form of an exponential divided by the square root of a determinant. Although compact, the resulting expression depends on the variables $z_i$ in a rather intricate manner, and it is instructive to organize this dependence as a power series in $z_i$. A systematic expansion yields
\be
    \tr\left[\hat G(x_1,z_1;v_1^{\rm\sst L},
    v_1^{\rm\sst R})\,
    \cdots\,
   \hat G(x_n,z_n;v_n^{\rm\sst L},v_n^{\rm\sst R})\right]
   =
\frac{\exp\left(\frac{\cL+i\,\cP}{2}\right)}
{\sqrt{x_{12}^2\,x_{23}^2\,\cdots\,x_{n1}^2}}\,,
\ee
in which the two distinct contributions $\cL$ and $\cP$ admit transparent diagrammatic interpretations. The first contribution, $\cL=\sum_{m=1}^{\infty}\cL^{(m)}$, originates from the expansion of the determinantal prefactor and is given by
\be
	\cL^{(m)}
	=(-i)^m
    \hspace{-30pt}
    \sum_{\substack{i_1,i_2,\ldots,i_m\\
   \text{with}\ i_{p+1}\,=\,i_{p}\pm1,\
   i_m=i_0}}
   \hspace{-30pt}
   z_{i_1}\,\cdots\,z_{i_m}\,
   \tfrac{1}{m}\,{\rm tr}\left(x^{-1}_{i_1\,i_1+1}\,x^{-1}_{i_2\,i_2+1}\,\cdots\,
   x^{-1}_{i_m\,i_m+1}\right).
\ee
Each $\cL^{(m)}$ thus collects all closed paths of length $m$ on the cyclic chain of sites, weighted by the corresponding product of inverse separations. The second contribution, $\cP=\sum_{i,j=1}^n\cP_{ij}^{(m)}$, arises from the Gaussian factor in the numerator and encodes the open paths connecting the polarization insertions:
\be
    \cP^{(m)}_{i_0i_m}=(-i)^m \hspace{-20pt}
    \sum_{\substack{
   i_0, i_1,\ldots,i_m \\
   \text{with}\ i_{p+1}\,=\,i_{p}\pm1}}
   \hspace{-20pt}
   z_{i_1}\,\cdots\,z_{i_m}
   \la v^{\sst\rm L}_{i_0}
   +v^{\sst\rm R}_{i_0+1}
   |\,x^{-1}_{i_0\,i_0+1}\,
   x^{-1}_{i_1\,i_1+1}\,
\cdots\,x^{-1}_{i_m\,i_{m}+1}\,|
v_{i_m}^{\sst\rm L}+v_{i_m+1}^{\sst\rm R}\ra\,.
\ee

\subsection{Boundary correlation functions}

Having assembled the bulk $n$-point traces in their generating form, we now extract the boundary correlation functions of the dual CFT by taking the limit $z_i \to 0$ and reading off the leading behavior. The discussion is organized according to the species of vertex operators inserted: we begin with the spinning sectors of both theories and then turn to the case in which pseudo-scalars are present.

\paragraph{Without pseudo-scalars.}
We first consider boundary traces built solely from the type-A spin-$s$ and scalar vertex operators, or from the type-B spin-$s$ vertex operators. As noted earlier, both the spin-$s$ vertex operator and the type-A scalar vertex operator scale as $z^{1+s}$ near the boundary, whereas the type-B pseudo-scalar exhibits a distinct scaling and will be treated separately below. In terms of the rescaled vertex operators \eqref{rescale A} and \eqref{rescale B}, the boundary limit therefore amounts to retaining the leading finite part of the trace as $z_i \to 0$.

At this order, only the $m=0$ contribution to the path sum survives, since every higher-$m$ term carries explicit powers of $z_i$ and accordingly vanishes in the boundary limit. The relevant piece is thus $\cP^{\sst (0)}$, and the corresponding boundary $n$-point traces can be evaluated in closed form. For the type-A theory, we obtain
\ba
    && \tr\left[
    \tfrac{1}{z_1}\,\hat V_{\rm A}\big(x_1,z_1;\tfrac{v_1}{\sqrt{z_1}}\big)\,
    \cdots\,
    \tfrac{1}{z_n}\,\hat V_{\rm A}\big(x_n,z_n;\tfrac{v_n}{\sqrt{z_n}}\big)
    \right]\nn
    &&
    =\,g^{n \phantom{g}}_{\rm HSG}\, \frac{
     \prod_{i=1}^n
    \exp\left(\tfrac{i}{2}\,Q_i\right)
     \prod_{i=1}^n
    \cos(P_{i\,i+1})}
    {\sqrt{x_{12}^2\,\cdots\,x_{n1}^2}}
    +\cO(z_i)\,,\label{general A trace}
\ea
where, for compactness, we have introduced the bilinears
\be
    Q_i= \la v_{i}|\left(
   x_{i\,i+1}^{-1}
   -
   x_{i\,i-1}^{-1}\right)|v_{i}\ra\,,
   \qquad
   P_{i\,i+1}=R_{i\,i+1}\,,
   \qquad
   R_{ij}=
   \la v_{i}|\,
   x^{-1}_{i\,i+1}\,\cdots\,
   x_{j-1\,j}^{-1}\,|v_{j}\ra\,.
\ee
The notation $P_{i\,i+1}$ is used to match the conventions of \cite{Didenko:2013bj}. Here $Q_i$ collects the diagonal terms localized at site $i$, while $R_{ij}$---defined for arbitrary $j$ in view of its later use---encodes the open path connecting the polarization spinors $v_i$ and $v_j$ through consecutive inverse separations. The cosines in \eqref{general A trace} arise from the even projection on the polarizations characteristic of the type-A theory. The analogous result for the type-B theory is obtained by replacing each cosine with a sine, reflecting the odd projection on the polarizations:
\ba
    && \tr\left[
    \tfrac{1}{z_1}\,\hat V_{\rm B}\big(x_1,z_1;\tfrac{v_1}{\sqrt{z_1}}\big)\,
    \cdots\,
    \tfrac{1}{z_n}\,\hat V_{\rm B}\big(x_n,z_n;\tfrac{v_n}{\sqrt{z_n}}\big)
    \right]\nn
    &&
   = i^n\,g^{n \phantom{g}}_{\rm HSG}\,\frac{
     \prod_{i=1}^n
    \exp\left(\tfrac{i}{2}\,Q_i\right)
     \prod_{i=1}^n
    i\,\sin(P_{i\,i+1})}
    {\sqrt{x_{12}^2\,\cdots\, x_{n1}^2}}
    +\cO(z_i)\,.\label{general B trace}
\ea
\paragraph{With pseudo-scalars.}

We now turn to traces involving pseudo-scalar insertions. Since the type-B pseudo-scalar vertex operator scales as $z^{2}$ rather than $z^{1}$, each pseudo-scalar insertion at site $i$ contributes an additional linear factor of $z_i$ in the rescaled trace. The boundary limit therefore retains those terms of the $z_i$ expansion that are linear in each $z_i$ associated with a pseudo-scalar site, while remaining at the leading order in the spinning sites.

To organize this combinatorics, we partition the pseudo-scalar insertions into $k$ maximal groups of consecutive sites. The $g$-th group occupies the sites from $s_g+1$ through $e_g-1$, so that $s_g$ and $e_g$ label the nearest spinning sites flanking the group. By construction, $s_g+1\le e_g-1$, the size of the $g$-th group is $n_g = e_g - s_g - 1$, and the total number of pseudo-scalar insertions is $\sum_{g=1}^{k}n_g$. The groups are ordered cyclically, with $e_g \le s_{g+1}$. With these conventions, the boundary trace evaluates to
\ba
    && \tr\left[
    \text{mixed insertions}
    \right] \nn
    && = g^{n \phantom{g}}_{\rm HSG}\,\frac{
    \prod_{g=1}^k\left[
    (-1)\,R_{s_g\,e_g}
     \prod_{j=e_g}^{s_{g+1}}
    \exp\left(\tfrac{i}{2}\,Q_j\right)
     \prod_{j=e_g}^{s_{g+1}-1}
    (-1)\,\sin(P_{j\,j+1})\right]}
    {\sqrt{x_{12}^2\,\cdots\, x_{n1}^2}}
    +\cO(z_i)\,.\quad
\ea 

\paragraph{Two-point functions.}

We now specialize the general formulas to the lowest correlators and compare them with the standard results of free vector models. For the type-A theory, the two-point function of the bilinear current $C^{\sst({\rm E})}$ takes the form 
\be
    \lim_{\{z_i\}\to 0}\left\la \tfrac{1}{z_1}\,C^{\sst ({\rm E})}\Big(x_1,z_1;\tfrac{v_1}{\sqrt{z_1}}\Big)\,
    \tfrac{1}{z_2}\,C^{\sst ({\rm E})}\Big(x_2,z_2;\tfrac{v_2}{\sqrt{z_2}}\Big)\right\ra^{\rm Type\,A}= \frac{\cos^2 (P_{12})}
    {x_{12}^2}\,,
\ee
while the type-B counterpart, governed by the odd projection, reads
\be
    \lim_{\{z_i\}\to 0}\left\la \tfrac{1}{z_1}\,C^{\sst ({\rm E})}\Big(x_1,z_1;\tfrac{v_1}{\sqrt{z_1}}\Big)\,
    \tfrac{1}{z_2}\,C^{\sst ({\rm E})}\Big(x_2,z_2;\tfrac{v_2}{\sqrt{z_2}}\Big)\right\ra^{\rm Type\,B}= -\frac{\sin^2 (P_{12})}
{x_{12}^2}\,.\label{type-B 2pt}
\ee
The minus sign in $-\sin^2(P_{12})$ may appear to violate positivity, but both $-\sin^2(P_{12})=(\cos(2P_{12})-1)/2$ and $\cos^2(P_{12})=(\cos(2P_{12})+1)/2$ have negative coefficients for odd spins upon expansion in spin, which is in fact consistent.\footnote{For the spin-1 example, with $V_1^{\a\b}=v_1^\a\,v_1^\b$ and $V_2^{\a\b}=v_2^\a\,v_2^\b$, one can show that $-(P_{12})^2$ takes the vector form
\be
    -(P_{12})^2 = 2\,V_1^\mu\left(\eta_{\m\n}\,x^2-2\,x_\mu\,x_\nu\right) V_2^\nu\,,
\ee
in the mostly plus signature. Upon Wick rotation, this becomes the positive-definite structure.}
The pseudo-scalar two-point function in the type-B theory is determined separately and yields
\be
    \lim_{\{z_i\}\to 0}\left\la \tfrac{1}{z_1^2}\,\Phi(x_1,z_1)\,
    \tfrac{1}{z_2^2}\,\Phi(x_2,z_2)\right\ra^{\rm Type\,B}=\frac{1}{(x^2_{12})^2}\,,\label{type-B 2pt phi}
\ee
which exhibits the conformal dimension expected of a parity-odd scalar in three dimensions.

\paragraph{Three-point functions.}

The three-point functions provide the first nontrivial test of the higher-spin structure. A general feature is that the full trace receives contributions from two cyclic orderings of the vertex-operator insertions, which differ by a permutation. Summing the two contributions converts each $e^{iQ_i/2}$ factor in \eqref{general A trace} and \eqref{general B trace} into either a cosine or a sine, depending on the relative sign of the two terms under the permutation. For the type-A theory, the two orderings add constructively, and we obtain
\ba
    &&\lim_{\{z_i\}\to 0}\left\la \tfrac{1}{z_1}\,C^{\sst ({\rm E})}\Big(x_1,z_1;\tfrac{v_1}{\sqrt{z_1}}\Big)\,
    \tfrac{1}{z_2}\,C^{\sst ({\rm E})}\Big(x_2,z_2;\tfrac{v_2}{\sqrt{z_2}}\Big)\,
    \tfrac{1}{z_3}\,C^{\sst ({\rm E})}\Big(x_3,z_3;
    \tfrac{v_3}{\sqrt{z_3}}\Big)\right\ra^{\rm Type\,A} \nn
    &&= 2\,g^{\phantom{g}}_{\rm HSG}\,\frac{\cos\left(\frac{Q_1+Q_2+Q_3}{2}\right)\cos (P_{12})\,\cos (P_{23})\,\cos (P_{31})}
    {\sqrt{x_{12}^2\,x_{23}^2\,x_{31}^2}}\,,\label{type-A 3pt}
\ea
in which the three cosines $\cos(P_{ij})$ correspond to the open paths along the edges of the triangle of insertion points, while $\cos((Q_1+Q_2+Q_3)/2)$ collects the local site contributions symmetrized over the two orderings. The analogous type-B result is obtained by trading every cosine for a sine,
\ba
    &&\lim_{\{z_i\}\to 0}\left\la \tfrac{1}{z_1}\,C^{\sst ({\rm E})}\Big(x_1,z_1;\tfrac{v_1}{\sqrt{z_1}}\Big)\,
    \tfrac{1}{z_2}\,C^{\sst ({\rm E})}\Big(x_2,z_2;\tfrac{v_2}{\sqrt{z_2}}\Big)\,
    \tfrac{1}{z_3}\,C^{\sst ({\rm E})}\Big(x_3,z_3;
    \tfrac{v_3}{\sqrt{z_3}}\Big)\right\ra^{\rm Type\,B} \nn
    &&= -2i\,g^{\phantom{g}}_{\rm HSG}\,\frac{\sin\left(\frac{Q_1+Q_2+Q_3}{2}\right)\sin (P_{12})\,\sin (P_{23})\,\sin (P_{31})}
    {\sqrt{x_{12}^2\,x_{23}^2\,x_{31}^2}}\,,\label{type-B 3pt}
\ea
Mixed correlators involving pseudo-scalar insertions interpolate between these two structures. With a single pseudo-scalar insertion,
\ba
    &&\lim_{\{z_i\}\to 0}\left\la \tfrac{1}{z_1}\,C^{\sst ({\rm E})}\Big(x_1,z_1;\tfrac{v_1}{\sqrt{z_1}}\Big)\,
    \tfrac{1}{z_2}\,C^{\sst ({\rm E})}\Big(x_2,z_2;\tfrac{v_2}{\sqrt{z_2}}\Big)\,
    \tfrac{1}{z_3^2}\,
\Phi(x_3,z_3)\right\ra^{\rm Type\,B} \nn
    &&= 2\,g^{\phantom{g}}_{\rm HSG}\,\frac{\cos\left(\frac{Q_1+Q_2}{2}\right) \sin(P_{12})\,R_{21}}{\sqrt{x_{12}^2\,x_{23}^2\,x_{31}^2}}\,,
\ea
the spinning bilinear at sites $1$ and $2$ produces the $\sin(P_{12})$ factor, while the pseudo-scalar at site $3$ contributes the open path $R_{21}$ traversing it. With two pseudo-scalar insertions, the structure further simplifies to
\be
    \lim_{\{z_i\}\to 0}\left\la
    \tfrac{1}{z_1}\,C^{\sst ({\rm E})}\Big(x_1,z_1;\tfrac{v_1}{\sqrt{z_1}}\Big)\,
    \tfrac{1}{z_2^2}\,\Phi(x_2,z_2)\,
    \tfrac{1}{z_3^2}\,
\Phi(x_3,z_3)\right\ra^{\rm Type\,B}=-2i\,g^{\phantom{g}}_{\rm HSG}\,
   \frac{\sin\Big(\frac{1}{2}Q_1\Big)\,R_{11}}{\sqrt{x_{12}^2\,x_{23}^2\,x_{31}^2}}\,,
\ee
with only a single spinning site contributing $\sin(Q_1/2)$ and the pseudo-scalar pair generating the closed path $R_{11}$. Finally, the purely pseudo-scalar three-point function vanishes,
\be
    \lim_{\{z_i\}\to 0}\left\la
    \tfrac{1}{z_1^2}\,\Phi(x_1,z_1)\,
    \tfrac{1}{z_2^2}\,\Phi(x_2,z_2)\,
    \tfrac{1}{z_3^2}\,
\Phi(x_3,z_3)\right\ra^{\rm Type\,B}=0\,,
\ee
since the two cyclic orderings contribute with opposite signs and cancel.

\subsection{Comparison with earlier works}

Having presented our results, we now place them in context and compare with several closely related lines of research. We first discuss the agreement of our boundary correlators with the standard CFT computations of higher-spin currents, then turn to the technical and conceptual parallels with the 
trace calculations in twisted-adjoint sector of Vasiliev equation, and finally examine the close kinship of our framework with the brane-parton proposal of Engquist and Sundell.

\paragraph{CFT results.}
Taken together, our results reproduce precisely the boundary correlators of the free boson and free fermion vector models---see, e.g., \cite{Giombi:2009wh,Giombi:2011rz,Sezgin:2017jgm}
and also
\cite{Didenko:2012tv,Colombo:2012jx,Didenko:2013bj}---thereby furnishing a nontrivial check of the duality at the level of $n$-point functions. The agreement may be traced back, at a more structural level, to a correspondence between the worldline vertex operators and the generating functions of higher-spin conserved currents in the dual CFTs. The generating functions of the type-A and type-B currents, together with that of the pseudo-scalar operator, were constructed in \cite{Sezgin:2017jgm} in the form
\ba
    && j_{\rm A}(\mathtt{x};v)=\cos(2e^{i\pi/4}\sqrt{U_1})\,
    \cos(2e^{i\pi/4}\sqrt{-U_2})\,\bar\phi_a(\mathtt{x}_1)\,\phi^a(\mathtt{x}_2)\,\big|_{\mathtt{x}_1=\mathtt{x}_2=\mathtt{x}}\,, \nn
    &&  j_{\rm B}(\mathtt{x};v)=
     \frac{1}{\sqrt{-U_1\,U_2}}\,\sin(2e^{i\pi/4}\sqrt{U_1})\,
    \sin(2e^{i\pi/4}\sqrt{-U_2})\,v^\a\,v^\b\,\bar\psi^a_\a(\mathtt{x}_1)\,\psi_{a\b}(\mathtt{x}_2)\,\big|_{\mathtt{x}_1=\mathtt{x}_2=\mathtt{x}}\,, \nn
    && \tilde j_{\rm B}(\mathtt{x})
    =\bar\psi^a_\a(\mathtt{x})\,\psi_a^\a(\mathtt{x})\,,
\ea
with
\be
U_1=\tfrac{1}{2}\,v^\a\,v^\b\,\partial^1_{\a\b}\,,
    \qquad
U_2=\tfrac{1}{2}\,v^\a\,v^\b\,\partial^2_{\a\b}\,.
\ee
Performing the on-shell Fourier transform of the boundary fields and substituting the twistor representation of the momenta, one finds that these generating functions coincide with the boundary limit of the spinning vertex operators \eqref{V type-A} and \eqref{V type-B}, as well as with the pseudo-scalar vertex operator \eqref{V type-B 0}. The worldline construction therefore provides a bulk realization of the very generating functions that organize the higher-spin currents on the boundary.

We emphasize that the radial dependence $\pm z_i\la \pi_i\,\pi_j\ra$ in each vertex operator gives rise, upon Gaussian integration, to a rich pattern of subleading contributions in the boundary expansion, so that the full set of parity-even and parity-odd boundary correlators of the dual CFTs is recovered, in a uniform manner, from a single bulk worldline computation.

Our position-space correlation functions can be readily translated to momentum space by Fourier-transforming with respect to each insertion point $x_i$. The integral over $x_i$ produces a delta function,
\be
   \delta^3\Big(\l_{i\a}\,\bar\l_{i\b}-\pi_{i\a}\,\pi_{i\b}\,
    +\pi_{(i+1)\a}\,\pi_{(i+1)\b}\Big)\,,
\ee
where $p_{i\,\a\b}=\l_{i\a}\bar\l_{i\b}$ denote the external momenta in spinor-helicity form. The resulting delta functions enforce a set of nontrivial algebraic relations between the spinor-helicity variables $(\l_i,\bar\l_i)$ and the worldline twistor variables $\pi_i$, which constrain the kinematic structure of the momentum-space correlators in a manner reminiscent of the constraints encountered in \cite{Gopakumar:2003ns}. In particular, projecting the polarization spinors onto $v_i=\l_i$ or $v_i=\bar\l_i$ reproduces the spinor-helicity setup employed in \cite{Jain:2021vrv,Bala:2025gmz,Ansari:2025fvi}, in which the parity-even and parity-odd CFT three-point functions admit especially compact expressions. A detailed analysis of how our worldline correlators map onto these spinor-helicity structures would be an interesting topic for future investigation.

\paragraph{Twisted-adjoint sector of Vasiliev equation.}
It is also instructive to draw an analogy between the worldline observables \eqref{HSG correlator} and the results of \cite{Colombo:2012jx, Didenko:2012tv,Didenko:2013bj} obtained within
the twisted-adjoint sector of Vasiliev theory. 
There, the correlation function is likewise expressed as a permutation sum of a trace,
\ba\label{object2}
    \tr\Big(B(X;\mathtt{x}_1,Z_1|Y)\star\kappa\star B(X,\mathtt{x}_2,Z_2|Y)\star\cdots \star
    B(X;\mathtt{x}_n,Z_n|Y)\star \kappa\Big)\,,
\ea
where $B$ and $\kappa$ 
 denote the Weyl zero-forms and the Kleinian operator, respectively, so that $B\star \kappa$ transforms in the adjoint representation of the higher-spin algebra. Following the interpretation of \cite{Colombo:2012jx,Didenko:2012tv,Didenko:2013bj},
 and the computational details of \cite{Didenko:2012tv,Didenko:2013bj},
 the zero-form $B$ is the \emph{classical solution} of the free twisted-adjoint equation corresponding to the bulk-to-boundary propagator, with an appropriate higher-spin dressing factor, given explicitly by
\be
    B(x,z;\mathtt{x}_i,\z_i,\bar\z_i |y,\bar y)
    = K(x,z,\mathtt{x}_i)\,
    e^{-i\,y_\a\,F^{\a\dot\a}(x,z,\mathtt{x}_i)\,
    \bar y_{\dot\a}}
    \left(e^{i(\zeta_i^\a\,y_\a+\theta)}
    +{\rm c.c}
    +{\st \binom{\z\to -\z}{\bar\z\to-\bar\z}}
    \right),
\ee
with the functions $K$ and $F^{\a\dot\a}$ defined by
\ba
     && K(x,z,\mathtt{x}_i)=
    \frac{z}{(x-\mathtt{x}_i)^2+z^2}\,,\nn
    && F^{\a\dot\a}(x,z,\mathtt{x}_i)
    =-\left(\frac{2z}{(x-\mathtt{x}_i)^2+z^2}\,(x-\mathtt{x}_i)^{\a\dot\a}
    +\frac{(x-\mathtt{x}_i)^2-z^2 }{(x-\mathtt{x}_i)^2+z^2}\,i\,\epsilon^{\a\dot\a}\right).
\ea
Here, $X=(x,z)$ denotes a bulk reference point and the $\mathtt{x}_i$ are boundary points, while $Z_i=(\zeta_i,\bar\zeta_i)$ are the bulk-transported boundary polarizations $v^\a_i$,
\be
    \zeta^\a_i=
    K(x,z,\mathtt{x}_i)\left(
    \frac{1}{\sqrt{z}}
    (x-\mathtt{x}_i)^{\a\b}
    -\sqrt{z}\,i\,\e^{\a\b}\right)
    v_{i\b}\,.
\ee
The star product and the trace are taken with respect to the $Y=(y,\bar y)$ oscillators.

Our construction bears a clear resemblance to this framework. Indeed, the $y$ and $\bar y$ oscillators can be mapped to our variables $(\xi,\pi)$ through
\be
    y_\a=\xi_\a-i\,\pi_\a\,,
    \qquad
    \bar y_{\dot\a}
    =\xi_\a+i\,\pi_\a\,,
    \label{relation twistors}
\ee
so that the Gaussian structure of $B$ corresponds to that of the wave function $\la\pi_1,\pi_2|\,C^{\sst\rm (E)}|\Omega\ra$, while the twisted field $B\star \kappa$ plays the same role within the trace as our vertex operator $\hat V$. Beyond these technical parallels, however, a sharp conceptual distinction remains. The construction of \cite{Colombo:2012jx,Didenko:2012tv,Didenko:2013bj} rests on the standard semi-classical picture of the bulk theory, with bulk-to-boundary propagators playing the central role; yet, the would-be integration over the bulk point $(x,z)$ does not arise in the usual manner, since the dependence on the bulk transverse coordinate $x$ decouples automatically, while $z$ is taken at some reference value. In our worldline framework, by contrast, no such bulk integration point appears at any stage: the boundary correlators are obtained as the boundary limit of bulk correlators, the latter being given directly by traces over the worldline Hilbert space, without invoking a bulk-to-boundary propagator at all.

\paragraph{Brane partons.}
While our work bears a strong technical resemblance to \cite{Colombo:2012jx,Didenko:2012tv,Didenko:2013bj}, at the conceptual level it is essentially aligned with the brane-parton model proposed by Engquist and Sundell in \cite{Engquist:2005yt}. A key technical difference, however, is that the construction of \cite{Engquist:2005yt} is formulated in AdS$_D$, where the worldline model---based on $\mathfrak{sp}(2,\mathbb{R})$ constraints---accommodates both positive- and negative-energy states, whereas the four-dimensional twistor model employed here describes only the positive-energy sector. Since the negative-energy states play a significant role in the constructions of \cite{Engquist:2005yt}, a direct comparison is not straightforward.

More specifically, the passage from the twisted-adjoint vertex operator $\Phi$---the analogue of our wave function $|\Psi\ra$---to the massless vertex operator $\cV$---the analogue of our vertex operator $\hat V$---is implemented in \cite{Engquist:2005yt} by left multiplication with the Klein operator, $\cV=\Phi\star \k$.\footnote{We note that in \cite{Engquist:2005yt,Colombo:2012jx}, the twisted-adjoint zero-forms are denoted by $\Phi$, while their adjoint counterparts are denoted by $\cV=\Phi\star\k$. This convention should not be confused with that of \cite{Didenko:2012tv,Didenko:2013bj}, where the same objects are denoted by $B$ and $\Phi$, respectively.}
Beyond the fact that these vertex operators take a relatively implicit form, in contrast to the explicit expressions obtained here, further differences arise whose origin is not entirely transparent. In our case, for instance, the corresponding passage is implemented by the map $\pi_2 \to i\,\pi_2$, which differs from the twist map of Vasiliev theory: ours flips the sign of \emph{all} $\mathfrak{so}(2,3)$ generators, whereas the twist map does so only for the translations. Since the generators act differently in the two cases, a more careful comparison is needed to clarify the origin of this apparent discrepancy.

%%%%%%%%%%%%%%%%%%%%%%%%%%%%%%%%%%%%%%%%%%%%%%%%%%%%%
\section{Discussions}\label{sec:5}

In this section, we offer a broader discussion of the framework developed in this paper. We first highlight the parallels with string theory and the proposal that a worldsheet may be regarded as a fabric woven from many worldlines. We then reformulate the worldline model as the edge mode of a Poisson sigma model (PSM), in which several features---fractional branes, loop expansion, and the unoriented projection---become considerably more transparent. We conclude with a number of future directions.

\subsection{Comparison with string theory}

Since our correlation functions retain their dependence on the bulk coordinate $z$, one might naively expect them to hold throughout the bulk of AdS$_4$, with $z$ treated as a free parameter. This expectation, however, is incompatible with the fact that each vertex operator satisfies the free equations of motion, just as the wave functions do in \eqref{Bargmann SD} and \eqref{Bargmann ASD}.
Moreover, in view of the absence of local observables in quantum gravity, correlation functions in HSG are meaningful only at the boundary, where the asymptotic states are effectively free.

It is helpful to draw an analogy with string theory to clarify the underlying principle of our worldline model. In string theory, the fundamental fields are the coordinate functions $X^\mu(\tau,\sigma)$ mapping the worldsheet into the target space, while the momentum $p_\mu$ enters as external data. Integrating out $X^\mu$ produces scattering amplitudes as functions of $p_\mu$.

In our worldline model, the corresponding role is played by the twistor variables $\xi^\a$ and $\pi_\a$, of which only $\pi_\a$ is directly related to the transverse momentum, $p_{\a\b}=\pi_\a\,\pi_\b$. The worldline model is therefore naturally viewed as describing dynamics in momentum space rather than in position space, with $\xi^\a$ playing the role of the variable conjugate to $\pi_\a$ in the first-order formulation.

With this identification in hand, the analogy proceeds as follows. A string state is a wave functional of $X^\mu(\tau_0,\sigma)$ at fixed $\tau_0$, whereas an HSG state is a wave function of $\bm\pi_\a(\t_0)$. In string theory, the momentum $p_\mu$ labels distinct plane waves from which vertex operators are constructed; in the worldline model, the role of $p_\mu$ is played by the spacetime point $(x^{\a\b},z)$, which labels distinct Dirac delta functions from which vertex operators are constructed. The crucial parallel concerns the regime of validity: in string theory, plane waves solve the free equations, but they remain useful even in the interacting theory once they are sent to infinity as \emph{in} or \emph{out} states. In precisely the same way, the delta functions in our worldline model solve the free equations, but they can be used in the interacting theory once the corresponding spacetime points are sent to the asymptotic boundary. This dictionary is summarized in Table~\ref{tab:comparison}.
\begin{table}[hhh]
\centering
\begin{tabular}{|l|l|}
\hline
String theory & HSG worldline theory \\
\hline \hline
Worldsheet parameters $\tau, \sigma$ &
Worldline parameter $\tau$ \\
Position variable $X^\mu(\tau,\sigma)$ &
Twistor variable $\bm\pi_\a(\tau)$ \\
String state $\Psi[X^\mu(\tau_0,\sigma)]$ at fixed $\tau_0$ &
HSG state $\Psi(\bm\pi_\a(\tau_0))$ at fixed $\tau_0$\\
Momentum $p_\mu$ &
AdS$_4$ point $(x^{\a\b},z)$  \\
Plane wave $e^{i\,p_\mu\,X^\mu(\tau,\sigma)}$ &
Delta function $\delta^2(\xi^\a(\tau)-x^{\a\b}\,\pi_\b(\t)\pm i z\,\pi^\a(\tau))$ \\
\hline
\end{tabular}
\caption{Dictionary between string theory and the HSG worldline model.}
\label{tab:comparison}
\end{table}

A more direct connection to string theory may be sought by pursuing the proposal of \cite{Engquist:2005yt} that a worldsheet can be viewed as a fabric woven from countless worldlines. Beyond the conceptual picture and the available circumstantial evidence, a concrete realization of this idea remains to be developed. In this direction, the key missing ingredient is a prescription for increasing the number of worldlines while keeping the light spectrum unchanged. An algebraic answer to this question has in fact already been provided in \cite{Vasiliev:2012tv, Vasiliev:2018zer}. The next step is then to endow the resulting collection of world-threads with a two-dimensional topology---that is, to specify the weaving pattern. In either case, this line of development calls for an enlargement of the target space, accompanied by eventual $\mathbb{Z}_2$ orbifoldings.

\subsection{Formulation by Poisson sigma model}

The worldline model employed in this paper is, strictly speaking, a double-line model that governs the free propagation of higher-spin fields; to accommodate interactions, however, we were led to split the double line into two single lines connected at vertices. This splitting prescription, while operationally effective, partially obscures the underlying identity of the worldline theory.
A more unified picture is obtained by recognizing the worldline as the boundary mode of a two-dimensional theory---specifically, the Poisson sigma model (PSM)
\be
    S_{\rm PSM}
    =\int_{\Sigma} \eta_A\wedge\dd Z^A+\frac{1}{2}\,\mu^{AB}(Z)\,\eta_A\wedge \eta_B\,,
\ee
a possibility already pointed out in \cite{Engquist:2005yt}. Here, $Z^A$ are zero-forms and $\eta_A$ are one-forms on the worldsheet $\Sigma$. From this perspective, our worldline model arises as the dynamics of the edge modes of the PSM, and the doubled-line structure acquires an unambiguous origin as the two sides of an open worldsheet.

To verify this reduction explicitly, we integrate out the Lagrange multiplier $\eta_{A\tau}$, which produces the constraint
\be
    \partial_\sigma Z^A
    +\mu^{AB}(Z)\,\eta_{A\sigma}=0\,.
\ee
In the \emph{symplectic} case, in which $\mu^{AB}$ admits an inverse $\omega_{AB}$, this constraint can be solved uniquely for $\eta_{A\sigma}$, and the action reduces to a closed two-form on the target space,
\be
   S_{\rm PSM}=\int_\Sigma  \frac{1}{2}\,\omega_{AB}(Z)\,\dd Z^A\wedge \dd Z^B
   =\int_{\partial \Sigma}
   \theta_A(Z)\,\dd Z^A\,.
\ee
The second equality, which follows from Stokes' theorem, makes manifest that the bulk action localizes onto its boundary. Choosing the symplectic form to be the constant one introduced in \eqref{symplectic form}, $\omega_{AB}(Z)=\Omega_{AB}$, the symplectic potential becomes simply $\theta_A(Z)=\Omega_{AB}\,Z^B$, and the boundary action reproduces precisely the worldline theory studied in the preceding sections.

This equivalence holds at the level of the classical action. Whether it persists at the quantum level might be a more delicate question, and we do not attempt to settle it here. Proceeding nonetheless on the assumption that the worldline and PSM descriptions remain equivalent upon quantization, we may import the conceptual toolkit of open-string theory: vertex operators are inserted exclusively on the boundary; the endpoints of the open string---identified with the edges of the worldsheet that constitute the worldline---may be dressed with Chan--Paton factors; and the full machinery of worldsheet topology becomes available for organizing perturbation theory. For supersymmetric generalizations of the PSM developed along related lines, we refer the reader to \cite{Ikeda:1993fh,Arias:2016agc,Basile:2025kib}.

Even where the two pictures may eventually turn out to agree, the PSM perspective often renders certain features considerably more transparent than the worldline description does. A case in point is provided by fractional branes. Since the target twistor space of the PSM is a $\mathbb{Z}_2$ orbifold, it inherits a symplectic singularity---a conical singularity originating from the reflection symmetry. One may then contemplate configurations in which the boundary of the worldsheet is anchored to this singularity, in direct analogy with fractional branes located at orbifold fixed points. Nontrivial contributions of this type arise only in the multi-line setting: even for an $N$-line configuration, the availability of fractional-brane endpoints permits effective spectra built out of fewer than $N$ singletons. This mechanism plays a crucial role in preserving the massless spectrum of HSG, which would otherwise be jeopardized by the proliferation of constituent singletons in the multi-line construction, and may be viewed as the geometric counterpart of the algebraic resolution proposed in \cite{Vasiliev:2012tv,Vasiliev:2018zer}.

%%%%%%%%%%%%%%%%%%%%%%%%%%%%%%%%%%%%%%%%%%%%
\subsection{Loop expansion}

A second arena in which the worldline and PSM pictures may diverge is the loop expansion. On the worldline side, the prescription \eqref{WL rule} admits no contribution that can be naturally interpreted as a loop, and accessing such contributions would require an extension of the prescription itself. On the PSM side, by contrast, a loop expansion arises naturally and in close analogy with that of open-string theory: tree-level amplitudes correspond to worldsheets of disk topology, while loop corrections originate from worldsheets of higher topological complexity---typified by the annulus or by the torus with a disk excised.

In string theory, the bookkeeping of these topological contributions is performed by the dilaton coupling to the Euler density, which assigns to each worldsheet a weight determined by its Euler characteristic. The PSM, however, admits no analogous coupling built from its own field content, and a built-in loop-counting parameter is therefore unavailable. In its place, we have invoked the freedom to introduce a coupling constant $g^{\phantom{g}}_{\rm HSG}$ as a proportionality factor relating the correlation functions of the higher-spin fields to those computed from the worldline or PSM. Should the PSM be extended to include loops, this prescription would have to be revised accordingly.

A salient feature of the comparison is that the two pictures, despite their classical equivalence, organize loop contributions in inequivalent ways. A loop diagram on the PSM side carries a genuine worldsheet quantum effect---a contribution to the path integral that has no counterpart on the worldline side. In particular, a mixed-insertion annulus diagram in the PSM reduces classically to two disconnected worldlines and should therefore be discarded from the worldline perspective. The interplay between these two organizing principles, and a systematic understanding of how PSM loops translate---or fail to translate---into the worldline framework, remains an interesting question for future investigation.

\subsection{Unoriented theory}

So far, the vertices we have considered connect worldlines either of the same direction and the same color, or of opposite directions and different colors via the map $Z^A \to i\, Z^A$. If, in addition, the swap of the two colors---equivalently, the exchange of the two lines composing the doubled worldline---is admitted as a symmetry, the physical content can be organized according to its representations.

The color label distinguishes the two lines of the double-line theory and is carried, in the twistor variables, by the two pairs $\pi_1$ and $\pi_2$. The line-swap operation therefore amounts to the dual parity map $\tilde{\mathsf{S}}$ \eqref{tilde S}, defined by $(\pi_1, \pi_2) \to (\pi_2, \pi_1)$. From the PSM perspective, this map has a direct geometric interpretation: it is the orientation reversal of the open worldsheet---the worldsheet parity transformation familiar from the construction of unoriented open strings. Imposing it as a symmetry is therefore the higher-spin counterpart of passing from oriented to unoriented strings. From the explicit forms of the wave functions \eqref{wave function SD}
and \eqref{wave function ASD}, one finds that the action of $\tilde{\mathsf{S}}$ on a spin-$s$ state is
\be
    \text{Type A}\,:\quad \tilde{\mathsf{S}}=(-1)^s\,,
    \hspace{50pt}
    \text{Type B}\,:\quad \tilde{\mathsf{S}}=-(-1)^s\,.
\ee
Therefore, the truncation to the minimal theory with only even spins amounts to retaining the even (odd) representation in the type-A (type-B) theory.

As in string theory, this $\mathbb{Z}_2$ structure may be combined with a Chan--Paton dressing $\lambda_a$ with $a=1,\ldots,M$, which promotes the gauge group of HSG to $U(M)$. The unoriented projection then reduces $U(M)$ to either $O(M)$ or $U\!Sp(M)$, depending on whether even or odd eigenvalues of $\tilde{\mathsf{S}}$ are retained---with the assignment opposite for type A and type B, owing to the relative sign in the two formulas above. The extra minus sign in the type-B case may be understood as a reflection of the fermionic origin of the dual CFT.

We note that, once loop contributions are taken into account, the unoriented projection requires the inclusion of additional, unorientable worldsheets, or the corresponding worldlines. At one loop, in particular, the annulus contribution must be supplemented by that of the M\"obius strip. In the case of the partition function without operator insertions, the M\"obius contribution appears to correspond to the second term in
\be
Z_{\rm unoriented}(\beta) = \tfrac{1}{2}\,Z^2(\beta) + \tfrac{1}{2}\,Z(2\beta)\,,
\ee
where $Z^2(\beta)$ denotes the partition function of the oriented HSG, built from a doubled-line loop, whereas $Z(2\beta)$ encodes the M\"obius contribution, which is effectively equivalent to a single-line loop traversed twice.

\subsection{Future directions}

The framework developed in this paper provides, in a fairly direct sense, an operator-level map between the free CFTs on the boundary of AdS and a candidate ``quantum gravity'' model in the bulk, the latter originating from the doubled-line worldline model \eqref{twistor action}. As noted earlier, the worldline vertex operators are placed in natural one-to-one correspondence with the boundary higher-spin currents, hinting at a realization of the AdS/CFT correspondence at a first-principles level. It would be valuable to extend this analysis to the quantum regime, where the strength of this correspondence can be assessed more definitively.

Several technical observations made in the course of this work suggest deeper structural relations. In particular, the passage from the wave function to the vertex operator on the worldline side bears a close formal similarity to the passage from the twisted-adjoint to the adjoint representation in Vasiliev theory. Combined with the explicit dictionary \eqref{relation twistors} between the two sets of oscillators, this analogy---together with the residual mismatches in how the symmetries are realized on each side---may eventually be sharpened into a precise structural relation. Establishing such a relation would shed substantial light on the interplay between the worldline (and PSM) picture and Vasiliev's formulation of HSG. The alternative worldline approaches of \cite{Fedoruk:2005np,Fedoruk:2006it}, which were designed precisely to reproduce equations of Vasiliev type, may provide useful guidance in this direction.

Pushing the analogy with string theory further, a particularly interesting line of research would be to investigate whether the worldline or PSM analogue of the string $\beta$-function calculation for vertex operators can yield the field equations of HSG, in the same spirit as the supergravity equations of motion are obtained in string theory. Such a derivation may produce equations for HSG in which the locality properties of the Vasiliev system become more transparent.

Following the spirit of Witten's open string field theory \cite{Witten:1985cc, Witten:1986gi}, whose Chern--Simons-like form is comparatively simple, one may hope to apply a parallel logic to HSG and extract a genuine action principle for the Vasiliev system. A specific obstruction one must overcome is the absence of explicit spacetime-coordinate dependence in the worldline and PSM actions, which raises both conceptual and technical subtleties. A possible point of entry is offered by the vertex operators themselves, which do carry bulk-point dependence and, in their operator form, localize on what is naturally interpreted as a twistor incidence relation. As remarked above, the appearance of the imaginary unit in this relation suggests that an analytic continuation is unavoidable. Even at the level of the worldline correlator computations, we have encountered delicate convergence issues for the Gaussian integrals at timelike-separated points; recovering the desired Feynman propagator appears to require passing to Euclidean signature, with the Wick rotation acting consistently on the reality condition of the twistor variables themselves.

Different choices of reality condition correspond to different real forms of the Howe-dual pair $\big(Sp(4,\mathbb{C}), O(2,\mathbb{C})\big)$; see Appendix \ref{sec:Howe} for a brief introduction.
The pair $\big(Sp(4,\mathbb{R}), O(2)\big)$ relevant for AdS$_4$ has been the focus of the present work; the most natural next candidate is the pair $\big(Sp(1,1), O^*(2)\big)$, since $\mathfrak{sp}(1,1)$ is isomorphic to the dS$_4$ isometry algebra $\mathfrak{so}(1,4)$. The results of this paper can therefore be transported to the de Sitter setting, where the underlying twistor action coincides with the one used here: both dual pairs sit inside the same $Sp(8,\mathbb{R})$ symmetry of the eight-dimensional phase space spanned by $\xi_i^\a$ and $\pi_{i\a}$. Although the twistor space is shared, the natural polarization chosen to represent (A)dS physics differs between the two signatures, and the relation between the corresponding Hilbert-space wave functions becomes nontrivial upon quantization. Studying the dS$_4$ case along these lines should clarify the role of the antipodal map and of the $\alpha$-vacua in the construction of the wave function. Moreover, since dS$_4$ possesses both a past and a future boundary, the boundary-limit structure is expected to be considerably richer than in AdS$_4$. 
Whereas the AdS$_4$ vertex operators reduce, in the boundary limit, to the generating functions of conserved currents in CFT$_3$, no comparable conventional CFT$_3$ exists on either of the de Sitter boundaries; the worldline construction may therefore provide concrete input to the putative dS$_4$/CFT$_3$ correspondence; 
see, e.g., \cite{Anninos:2011ui,David:2020ptn,Lang:2025rxt,Giombi:2026sqa} for some attempts in this direction, particularly in the context of higher spins.

A closely related extension is to four-dimensional flat space, where the chiral higher-spin theory---see, e.g., \cite{Metsaev:1991mt,Metsaev:1991nb,Ponomarev:2016lrm,Adamo:2016ple,Sharapov:2022faa,Neiman:2024vit}---has been the subject of intensive recent activity and has yielded a number of positive results on tractable higher-spin interactions. Since the free theory of massless higher-spin fields admits a parallel twistor construction in flat space, it is natural to conjecture that a suitable gluing prescription applies in that setting as well, tailored to the chiral structure. We expect that the worldline higher-spin model proposed here may help to elucidate how to move beyond the chiral sector, whose constraint, while of considerable interest, is perhaps unnecessarily severe; see, e.g., \cite{Adamo:2022lah,Serrani:2026dbs} for some recent attempts in this direction.

A further direction, touched upon above, concerns the deeper role of the PSM at the quantum level. The PSM carries a built-in gauge symmetry, and the associated BRST charge $Q$ may furnish the entry point for a higher-spin analogue of open string field theory. Within this enlarged framework, a variety of loop-level effects---including vacuum-energy computations \cite{Giombi:2013fka,Giombi:2014iua} (see also \cite{Pang:2016ofv,Giombi:2016pvg,
Bae:2016rgm,
Bae:2016hfy,Bae:2016xmv,Basile:2018acb})---could profitably be revisited from a microscopic worldsheet perspective.

Finally, we should acknowledge an important limitation of the present approach: it does not capture what is arguably the most fascinating corner of higher-spin holography, namely the $\theta$-dependence of HSG and its matching to the Chern--Simons-coupled vector models on the CFT side \cite{Giombi:2011kc,Giombi:2012ms,Aharony:2012nh,Karch:2016sxi}. Vasiliev's equations themselves do encode this nontrivial relation for a certain class of three-point functions \cite{Giombi:2009wh,Giombi:2010vg, Giombi:2012ms, Sezgin:2017jgm}; by contrast, the Vasiliev-based methods of \cite{Colombo:2012jx,Didenko:2012tv,Didenko:2013bj}, as well as the worldline approach developed here, presently fall short of reproducing it. A speculative but appealing possibility is that the worldline theory, or the PSM in its multi-line extension, admits a deformation associated with discrete symmetries that would correspond, on the boundary, to turning on the Chern--Simons coupling. In this scenario, one might further envision a PSM containing not only open strings but also closed strings, whose boundary limit would generate the CFT operators built from the curvatures of the Chern--Simons gauge fields.

\section*{Acknowledgement}

M.C. and T.T. thank the Korea Institute for Advanced Study (KIAS) for its hospitality and for providing an environment conducive to various discussions during the course of this project. The work of M.C., E.J., and T.T. is supported by the National Research Foundation of Korea (NRF) grant funded by the Korean government (MSIT) (RS-2025-00564305). The work of T.O. is supported by KIAS Individual Grant (QP099501) via the Quantum Universe Center at the Korea Institute for Advanced Study. T.T. was also supported by the Young Scientist Training (YST) program at the Asia Pacific Center of Theoretical Physics (APCTP), through the Science and Technology Promotion Fund and Lottery Fund of the Korean Government, as well as by the Korean local governments---Gyeongsangbuk-do Province and Pohang City---during the initial stage of this work.

\appendix

\section{Vector/spinor dictionary}\label{sec: spinor convention}

The isomorphism $\mathrm{Spin}(1,2)\simeq Sp(2,\mathbb{R})$ allows three-dimensional vectorial objects to be expressed in spinorial form, and vice versa. In particular, the map to a 3-vector $V^\mu$ 
from the corresponding $Sp(2,\mathbb{R})$ tensor $V^{\a\b}$ is implemented with the help of the $\mathfrak{sp}(2,\mathbb{R})$ generators $\tau^\mu_{\a\b}$ as
\be
  V^\mu=-\frac12\,(\tau^\mu)_{\a\b}\,  V^{\a\b}\,,
\ee
where the $\tau^\mu$ matrices have the explicit realization
\be
    (\tau^0)_{\a\b}=
    \begin{pmatrix}
        1&0\\
        0&1
    \end{pmatrix}, \qquad
    (\tau^1)_{\a\b}=
    \begin{pmatrix}
        0&1\\
        1&0
    \end{pmatrix}, \qquad
    (\tau^2)_{\a\b}=
    \begin{pmatrix}
        1&0\\
        0&-1
    \end{pmatrix}.
\ee
Spinor indices are raised and lowered by the $\mathfrak{sp}(2,\mathbb{R})$-invariant tensor
\be
	  \e^{\a\b}= \e_{\a\b}=
    \begin{pmatrix}
        0&1\\
        -1&0
    \end{pmatrix},
\ee
according to
\be
    v^\a=\e^{\a\b}\,v_\b\,, \qquad
	v_\b=v^\a \e_{\a\b}\,, \qquad
	\e^{\a\g}\e_{\b\g}=\d^\a{}_\b\,,
    \qquad
   \la v\,w\ra= v_\a\,w^\a\,.
\ee
Useful relations for translating between the vectorial and spinorial descriptions of Lorentz-invariant objects are
\be\label{vector spinor dictionary}
    \eta^{\mu\nu}=-\frac12\,(\tau^\mu)_{\a\b}\,(\tau^\nu)^{\a\b}\,,\qquad V_\mu\,W^\mu=-\frac{1}{2}\,V_{\a\b}\,W^{\a\b}\,,
\ee
from which it follows that $V^\mu\,V_\mu = -\det V_{\a\b}$. Throughout the paper, we use the three-dimensional metric $\eta_{\mu\nu}={\rm diag}(-,+,+)$. For matrix operations we adopt the shorthand
\be \label{shortened}
    \la v\,\rvert\, V\,\rvert\,w\ra = v^\a\,V_{\a\b}\,w^\b\,,\qquad
    (V\,W)_{\a\d}=V_{\a\b}\,\e^{\b\g}\,W_{\g\d}\,,\qquad
    {\rm tr}(V) = V^\a{}_\a\,.
\ee

%%%%%%%%%%%%%%%%%%%%%%%%%%%%%%%%%%%%%%%%%%%%%%%%%%%%%%%%%%%%%%%%%%%%%%%%%%%%%%%%%%%%%%%
\section{Howe duality and the Flato--Fronsdal theorem}\label{sec:Howe}

Quite generally, the (reductive) dual pair correspondence \`a la Howe \cite{Howe:1989nad,Howe1989ii} states that the metaplectic representation $\cW_N$ of $Sp(2N,\mathbb{R})$ decomposes into irreps of a reductive subgroup $G\times \tilde G$ with multiplicity one,
\be
	\cW_N=\bigoplus_{\l}\, \cV^{G}_{\l}\otimes \cV^{\tilde G}_{\theta(\l)}\,,
\ee
where $G$ and $\tilde G$ are mutual centralizers. The $\tilde G$-representations $\cV^{\tilde G}_{\theta(\l)}$ appearing in this decomposition are in one-to-one correspondence with the $G$-representations $\cV^G_\l$ via the bijective map $\theta:\cV_\l^G\mapsto \cV_{\theta(\l)}^{\tilde G}$, so that specifying a representation of $\tilde G$ in $\cW_N$ is equivalent to specifying the corresponding representation of $G$. See \cite{Basile:2020gqi} for a review of the reductive dual pair correspondence.

In our context, we consider the $\cW_4$ representation of $Sp(8,\mathbb{R})$ associated with the reductive dual pair $\big(Sp(4,\mathbb{R}),\,O(2)\big)$. The phase space of the worldline action \eqref{twistor action}, parametrized by $(\bm\xi^\a,\bm\pi_\a)$, realizes $\cW_4$ as a representation of the AdS$_4$ algebra $\mathfrak{sp}(4,\mathbb{R})$ \eqref{sp4}. This algebra commutes with the spin operator \eqref{spin generator}, which generates the dual $\mathfrak{o}(2)$ algebra. By Howe's theorem, $\cW_4$ decomposes as
\be
	\cW_4=
	\Big((1,1)_{O(2)}\otimes \cD(2,\tilde 0)\Big)\oplus
	\bigoplus_{n=0}^\infty\,\Big((n)_{O(2)} \otimes \cD(\tfrac{n}{2}+1,\tfrac{n}{2})\Big)\,.   \label{W4}
\ee
Here, $\cD(2,\tilde 0)$ and $\cD(s+1,s)$ with $s\in \tfrac{1}{2}\mathbb{N}$ denote, respectively, the pseudo-scalar and massless spin-$s$ representations of the AdS$_4$ algebra. On the $O(2)$ side, the representations are labeled by traceless Young diagrams $(n)$ with $n=0,1,2,\ldots$, together with the pseudo-scalar Young diagram $(1,1)$. The $(n)_{O(2)}$ with $n\ge 1$ are doublets, while $(0)_{O(2)}$ and $(1,1)_{O(2)}$ are singlets. This is to be contrasted with the irreps of $U(1)$, which are all singlets: each $(n)_{O(2)}$ with $n\in \mathbb{N}_{>0}$ decomposes into $(+n)_{U(1)}\oplus (-n)_{U(1)}$.

This dual pair correspondence admits a particularly transparent specialization, the celebrated Flato--Fronsdal theorem \cite{Flato:1978qz}, via the seesaw dual pair \cite{Kudla1986,Kudla1996}
\be
\parbox{180pt}{
\begin{tikzpicture}
\draw [<->] (0,0) -- (1,0.8);
\draw [<->] (0,0.8) -- (1,0);
\node at (-1.2,1.2) {$O(2)$};
\node at (-1.2,0.4) {$\cup$};
\node at (2.2,-0.4) {$Sp(4,\mathbb{R})$};
\node at (-1.2,-0.4) {$O(1)\times O(1)$};
\node at (2.2,0.4) {$\cup$};
\node at (2.2,1.2) {$Sp(4,\mathbb{R})\times Sp(4,\mathbb{R})$};
\end{tikzpicture}}\,.  \label{seesaw O Sp}
\ee
On the left-hand side, restricting $O(2)$ to its subgroup $O(1)\times O(1)$, the singlet representations $(0)_{O(2)}$ and $(1,1)_{O(2)}$ decompose as
\bs
\begin{align}
 (0)_{O(2)}\Big|_{O(1)\times O(1)}&=(+)_{O(1)} \otimes (+)_{O(1)}\,,\\
	(1,1)_{O(2)}\Big|_{O(1)\times O(1)}&=(-)_{O(1)} \otimes (-)_{O(1)}\,.
\end{align}
\es
The reduced group $O(1)\times O(1)\simeq \mathbb{Z}_2\times \mathbb{Z}_2$ acts on the worldline variables as in \eqref{tilde P}, with $\tilde{\mathsf{P}}_1$ and $\tilde{\mathsf{P}}_2$ generating the left and right $\mathbb{Z}_2$ factors, respectively. Consequently, the $(+)_{O(1)}$ and $(-)_{O(1)}$ representations correspond to even and odd functions of $\pi_{I\a}$, respectively.

On the right-hand side of \eqref{seesaw O Sp}, the natural dual partners of $(\pm)_{O(1)}$ are the Rac and Di representations---namely, the 3d free boson $\cD(\tfrac{1}{2},0)$ and the 3d free fermion $\cD(1,\tfrac{1}{2})$. The pair $\big(O(1), Sp(4,\mathbb{R})\big)$ indeed forms a dual pair, giving the decomposition
\be
	\cW_2=\Big((+)_{O(1)}\otimes \cD(\tfrac{1}{2},0)\Big)
	\oplus \Big( (-)_{O(1)}\otimes \cD(1, \tfrac{1}{2}) \Big),
\ee
where
\be
	\cD(\tfrac{1}{2},0)={\rm FB}={\rm Rac}\,,
	\qquad
	\cD(1,\tfrac{1}{2})={\rm FF}={\rm Di}\,.
\ee
Turning to the higher-spin sector, the traceless Young diagrams $(n)_{O(2)}$ with $n>0$ decompose under $O(1)\times O(1)$ as
\be
	(n)_{O(2)}\Big|_{O(1)\times O(1)}=
	\Big((+)_{O(1)} \otimes ((-)^n)_{O(1)}\Big)
	\oplus
	\Big((-)_{O(1)} \otimes ((-)^{n+1})_{O(1)}\Big)\,.\label{eq:isomorphism}
\ee
Combining the scalar sector with the higher-spin sectors, we recover the Flato--Fronsdal theorem in the form
\bs\label{Flato Fronsdal Howe}
\begin{align}
	{\rm FB}\otimes {\rm FB}&=
	\Big((+)_{O(1)}\otimes (+)_{O(1)}\Big)\otimes
	\bigoplus_{n=0}^\infty \cD(n+1,n)\,,\\
	{\rm FF}\otimes {\rm FF}&=
	\Big((-)_{O(1)}\otimes (-)_{O(1)}\Big)\otimes
	\bigg(\cD(2,\tilde 0)\oplus \bigoplus_{n=1}^\infty \cD(n+1,n)\bigg),\\
	{\rm FB}\otimes {\rm FF}&=
	\Big((+)_{O(1)}\otimes (-)_{O(1)}\Big)\otimes
	\bigoplus_{n=0}^\infty \cD(n+\tfrac{3}{2},n+\tfrac{1}{2})\,,\\
	{\rm FF}\otimes {\rm FB}&=
	\Big((-)_{O(1)}\otimes (+)_{O(1)}\Big)\otimes
	\bigoplus_{n=0}^\infty \cD(n+\tfrac{3}{2},n+\tfrac{1}{2})\,.
\end{align}
\es
The $O(1)\times O(1)$ representations appearing here are all singlets, and simply record whether the $Sp(4,\mathbb{R})$ irreps are realized as even or odd functions of $\pi_{1\a}$ and $\pi_{2\a}$.

\section{Gaussian integral}\label{sec: Gaussian integral}

The traces of the form \eqref{start off Gaussian} in the main text can be computed by means of the matrix Gaussian integral identity
\be\label{Gaussian master}
	 I=\int \frac{\dd^{2n}\Pi}{(2\pi)^n}\,
	\exp\left(-\frac{i}{2}\,\Pi\cdot(A-B)\,\Pi+ \Pi\cdot W \right)
	=
	\frac{\exp\left(\frac{i}{2}\,W\cdot (A-B)^{-1}W\right)}{\sqrt{(-1)^n\det(A-B)}}\,.
\ee
Here, $A$ and $B$ are $2n\times 2n$ matrices,
\be
	A=\begin{pmatrix}
	a_{1} &   &    &  &  \\
	                     & a_2 &  &  & \\
	                     &  & \ddots &  &  \\
	                     & &   & a_{n-1} &  \\
	                      &  &  &  & a_{n}
	                     \end{pmatrix},
	                     \qquad
	B=\begin{pmatrix}
	 & b_{12}  &   &  & b_{1n} \\
	                   b_{21}  &  &  & &  \\
	                    &   & \ddots &  &  \\
	                    &  &   &  & b_{n-1\,n}\\
	                    b_{n1} & & & b_{n\,n-1} &
	                     \end{pmatrix},
\ee
with $2\times 2$ submatrices $a_i$ and $b_{ij}$. The indices $i,j$ are defined modulo $n$, and $b_{ij}^t=b_{ji}$ are defined only for adjacent $i$ and $j$. Likewise, $\Pi$ and $W$ are $2n$-component vectors,
\be
	\Pi=\begin{pmatrix} \pi_{1}
    \\\pi_2\\\vdots \\ \pi_n \end{pmatrix},
	\qquad
	W=\begin{pmatrix} w_1 \\ w_2\\\vdots \\ w_n \end{pmatrix},
\ee
with two-component subvectors $\pi_i$ and $w_i$. To extract the structure of \eqref{Gaussian master} in a useful form, we expand the result in powers of $B$, proceeding in two steps.

First, the exponent in the numerator of \eqref{Gaussian master} can be expanded as
\be
     \cP=W\cdot (A-B)^{-1}\,W =  \sum_{m=0}^{\infty}
   W \cdot A^{-1}\,(B\,A^{-1})^m\,W
	=\sum_{m=0}^\infty \sum_{i,j=1}^n \cP_{ij}^{(m)}\,,
\ee
where the $\cP^{(m)}_{ij}$ are given combinatorially by
\be
    \cP^{(m)}_{i_0i_m}= \sum_{\substack{\text{all length-$m$ paths}\\
   i_0\,\to\, i_1\,\to\, \cdots\, \to\, i_m\, \\
   \text{with}\ i_{p+1}\,=\,i_{p}\pm1}}
   \la w_{i_0}|\,a^{-1}_{i_0}\,b_{i_0i_1}\,a^{-1}_{i_1}\,
   b_{i_1i_2}\,\cdots\,
   a^{-1}_{i_m}\,|w_{i_m}\ra\,.
\ee
The intermediate indices $i_k$ are just labels of the same kind as $i$ and $j$.

Second, the determinant in the denominator of \eqref{Gaussian master} can be expanded as
\be
	\det(A-B)=\det A\,\exp(-\cL)\,,
\ee
in which the dominant contribution arises from
\be \label{correl function dominant}
	\det A= \det a_1\,\det a_2 \,\cdots\,\det a_{n}\,,
\ee
and $\cL$ is given by the series
\be
    \cL=\sum_{m=1}^\infty \cL^{(m)}\,,
\ee
with
\be
	\cL^{(m)}=\frac{1}{m}\,{\rm tr}[(B\,A^{-1})^m]
	=\sum_{\substack{\text{all length-$m$ closed paths}\\
   i_0\,\to\, i_1\,\to\ \cdots\ \to\, i_m\,=\,i_0\\
   \text{with}\ i_{p+1}\,=\,i_{p}\pm1}}
   \frac{1}{m}\,{\rm tr}\left(b_{i_0i_1}\,a^{-1}_{i_1}\,b_{i_1i_2}\,\cdots\,
   a^{-1}_{i_m}\right).
\ee
In summary, the Gaussian integral reads
\be
   I=
    \frac{\exp(\cM)}{\sqrt{(-\det a_1)(-\det a_2)\cdots (-\det a_n)}}\,,
\ee
with
\be
    \cM=\frac{1}{2}\left(\cL+i\,\cP\right)
    =\frac{1}{2}\sum_{m=0}^\infty
    \left(\cL^{(m)}+i\,\cP^{(m)}\right),
\ee
where $m$ counts the total power of $b_{ij}$, and $\cL^{(0)}=0$.

If we restrict to contributions linear in each $b_{ij}$, these correspond to self-avoiding (SA) paths, for which the surviving contributions are
\be
    \begin{array}{l}
    \left[\cP^{(m)}_{i\,i+m}\right]_{\rm SA}
     =
    \la w_{i}|\,a^{-1}_{i}\,b_{i\,i+ 1}\,a^{-1}_{i+1}\,
   b_{i+ 1\,i+ 2}\,\cdots\,
   a^{-1}_{i+ m}\,|w_{i+m}\ra \\
   \hspace{70pt} +\,\delta_{n>2}\,\delta^{\, m}_{n/2}\,\la w_{i}|
   \,a^{-1}_{i}\,b_{i\,i-1}\,a^{-1}_{i-1}\,
   b_{i-1\,i-2}\,\cdots\,a_{i-m}^{-1}\,|w_{i-m}\ra
   \medskip\\
    \Big[\cP^{(m)}_{i+m\,i}\Big]_{\rm SA}= \Big[\cP^{(m)}_{i\,i+m}\Big]_{\rm SA}
   \end{array}
   \qquad [m\le n]\,,
\ee
where $\delta$ denotes the Kronecker delta symbol, and
\be
	 \left[\cL^{(m)}\right]_{\rm SA}
	=\d_n^m\,2\,
   {\rm tr}\left(b_{n1}\,a^{-1}_{1}\,b_{12}\,\cdots\,
   a^{-1}_{n}\right),
\ee
where $\left[\cP^{(m)}_{i\,i\pm m}\right]_{\rm SA}$ corresponds to the monotonic forward/backward path, with an opposite-orientation contribution arising only at $n=2m>2$, and the overall factor of $2$ in $\left[\cL^{(m)}\right]_{\rm SA}$ accounts for the clockwise and anti-clockwise cycles, each admitting $n$ possible starting positions.

We now apply these results to the computation of the $n$-point trace of the rescaled vertex operators, for which the matrices $a_i$ and $b_{ij}$ and the vectors $w_i$ take the form
\be
    a_i=x_{i\,i+1}\,,\qquad
    b_{i\,i-1}=-b_{i-1\,i}=i\,z_i\,\e\,,\qquad
    w_i=v_{i}^{\sst\rm L}+v_{i+1}^{\sst\rm R}\,,
\ee
where the components $x_{i\,i+1}$, $\e$, and $v_i$ all carry upper indices, namely $(x_{ij})^{\a\b}$, $\e^{\a\b}$, and $v_i^\a$. The Gaussian integrals are then projected appropriately according to whether the correlator is of type A or type B.

The leading contribution in the $z_i$ expansion arises entirely from
\be
    \cP^{(0)}_{ii}
    =
     \la v_{i}^{\sst\rm L}|\,
   x_{i\,i+1}^{-1}\,|v_{i}^{\sst\rm L}\ra
   +
    \la v_{i+1}^{\sst\rm R}|\,
   x_{i\,i+1}^{-1}\,|v_{i+1}^{\sst\rm R}\ra
   +2\,\la v_{i}^{\sst\rm L}|\,
   x_{i\,i+1}^{-1}\,|v_{i+1}^{\sst\rm R}\ra\,.
\ee
The only subleading contribution comes from the pseudo-scalar, and the sub-subleading contributions vanish in the boundary limit. It therefore suffices to retain only
\ba
    &&
    \left[\cP^{(m)}_{i\,i+m}\right]_{\rm SA} =
    (-i)^m z_{i+1} \cdots z_{i+m}
   \la v_{i}^{\sst\rm L}+v_{i+1}^{\sst\rm R}|\,
   x_{i\,i+1}^{-1}\,\cdots\,x_{i+m\,i+m+1}^{-1}\,|v_{i+m}^{\sst\rm L}+v_{i+m+1}^{\sst\rm R}\ra
   \nn
   &&\qquad +\,\d^m_{n/2} (-i)^m z_{i} \cdots z_{i-m+1}
   \la v_{i}^{\sst\rm L}+v_{i+1}^{\sst\rm R}|\,x_{i+1\,i}^{-1}\,\cdots\,x_{i-m+1\,i-m}^{-1}\,|v_{i-m}^{\sst\rm L}+v_{i-m+1}^{\sst\rm R}\ra\,,
\ea
for $1\le m\le n$, and
\be
	\left[\cL^{(m)}\right]_{\rm SA}
	=\d_n^m\,2\,(-i)^n\,
    z_1\,\cdots\,z_n\,
    {\rm tr}\left(x_{12}^{-1}\,x_{23}^{-1}\,\cdots\,
   x_{n-1\,n}^{-1}\,x_{n1}^{-1}\right).
\ee
Moreover, since only the pseudo-scalar contributes at subleading order in $z_i$, no polarization $v_i$ may appear. We refer to this case as \emph{strong self-avoidance} (SSA), for which only one of the four terms above is relevant:
\ba
    \left[\cP^{(m)}_{i\,i+m}\right]_{\rm SSA} \eq 
   (-i)^m\,z_{i+1}\,\cdots\, z_{i+m}\,
    \la v_{i}^{\sst\rm L}|\,
x_{i\,i+1}^{-1}\,\cdots\,x_{i+m\,i+m+1}^{-1}\,|v_{i+m+1}^{\sst\rm R}\ra\nn
   && +\,\d^{m}_{n/2}\,(-i)^m\,
   z_{i-m+1}\,\cdots\, z_{i}\,
    \la v_{i-m}^{\sst\rm L}|\,
x_{i-m\,i-m+1}^{-1}\,\cdots\,x_{i\,i+1}^{-1}\,
|v_{i+1}^{\sst\rm R}\ra,
\ea
for $1\le m\le n$. The relevant part of the exponent is then
\ba
    &&\left[\cM\right]_{\rm SSA}=
    (-i)^n\,
    z_1\,\cdots\,z_n\,
    {\rm tr}\left(x_{12}^{-1}\,x_{23}^{-1}\,\cdots\,
   x_{n-1\,n}^{-1}\,x_{n1}^{-1}\right)
   +\frac{i}{2}\sum_{i=1}^n
   \la v_{i}|\left(
   x_{i\,i+1}^{-1}
   +
   x_{i-1\,i}^{-1}\right)|v_{i}\ra
   \nn
   &&\ +\,i\,
   \sum_{m=0}^n\sum_{i=1}^n
   (-i)^m\,\big(1+\d^m_{n/2}\big)\,z_{i+1}\,\cdots\, z_{i+m}\,
    \la v_{i}^{\sst\rm L}|\,
x_{i\,i+1}^{-1}\,\cdots\,x_{i+m\,i+m+1}^{-1}\,|v_{i+m+1}^{\sst\rm R}\ra\,,
\ea
where the quadratic terms in $v_{i}^{\rm\sst L}$ and $v_{i}^{\rm\sst R}$ have already been replaced by $v_i$, as they are irrelevant under the even/odd projection. Exponentiating the above and projecting onto the SSA sector, we obtain the generating function of strong self-avoiding disconnected paths,
\ba
    \left[\exp\left(\cM\right)\right]_{\rm SSA}
    \eq
    \prod_{i=1}^n \exp\left(
    \tfrac{i}{2} \la v_{i}|\left(
   x_{i\,i+1}^{-1}
   +
   x_{i-1\,i}^{-1}\right)|v_{i}\ra\right)
    \prod_{i=1}^n
    \exp\left(
    i\la v_{i}^{\sst\rm L}|\,
x_{i\,i+1}^{-1}\,|v_{i+1}^{\sst\rm R}\ra
    \right)\times
    \nn
    && \times\,
    \prod_{k_1=0}^1\cdots \prod_{k_n=0}^1
    (-i\,z_1)^{1-k_1}\,\cdots\,(-i\,z_n)^{1-k_n}\,\cN_{k_1\cdots k_n}\,,
\ea
where the coefficients $\cN_{k_1\cdots k_n}$ are defined by
\ba
    && \cN_{1\cdots 1}=1\,,
    \qquad \cN_{0\cdots 0}= {\rm tr}\left(x_{12}^{-1}\,x_{23}^{-1}\,\cdots\,
   x_{n-1\,n}^{-1}\,x_{n1}^{-1}\right), \nn
    && \cN_{\text{---}\,\underset{{i}}{1}0\cdots 0 \underset{{j}}{1}\,\text{---}}
    =\cN_{\text{---}\,\underset{{i}}{1}\cdots\cdots  \underset{ {j}}{1}\,\text{---}}
    \times
    i\,\left(1+\delta^{[{j}-{i}]_{n}-1}_{n/2}\right)\la v_{{i}}^{\sst\rm L}|\,
x_{{i}\,{i}+1}^{-1}\,\cdots\,x_{{j}-1\,{j}}^{-1}\,|v_{{j}}^{\sst\rm R}\ra\,,
    \label{Mssa}
\ea
where the indices $i$ with $k_i=0$ mark the locations of the pseudo-scalar insertions, and $[j-i]_n=(j-i\,\mathrm{mod}\, n)$.

%%%%%%%%%%%%%%%%%%%%%%%%%%%%%%%%%%%%%%%%%%%%%%%%%%%%%%%%%%%%%%%%%%%%%%%%

\bibliographystyle{JHEP}
\bibliography{biblio}

\end{document}